\documentclass[%
 reprint,
superscriptaddress,
 amsmath,amssymb,
 aps,
prb,
]{revtex4-2}

\usepackage{color}
\usepackage{graphicx}
\usepackage{dcolumn}
\usepackage{bm}
\usepackage{hyperref}
\hypersetup{colorlinks=true,linkcolor=blue,citecolor=blue,urlcolor=blue}

\usepackage[T1]{fontenc} 
\usepackage{newtxtext}   
\usepackage{newtxmath}   
\usepackage{upgreek}
\usepackage{mhchem}

\begin{document}

\preprint{APS/123-QED}

\title{Probing critical spin fluctuations with a composite magnetoelectric method: A case study on a Kitaev spin liquid candidate \ce{Na3Co2SbO6} }  

\author{Xinrun Mi}
\thanks{These authors contributed equally to this work. }
\affiliation{College of Physics \& Center of Quantum Materials and Devices, Chongqing University, Chongqing 401331, China}

\author{Xintong Li}
\thanks{These authors contributed equally to this work. }
\affiliation{International Center for Quantum Materials, School of Physics, Peking University, Beijing 100871, China}
\affiliation{National Lab for Superconductivity, Beijing National Laboratory for Condensed Matter Physics, Institue of Physics, Chinese Academy of Sciences, Beijing 100190, China }

\author{Long Zhang}
\thanks{These authors contributed equally to this work. }
\affiliation{College of Physics \& Center of Quantum Materials and Devices, Chongqing University, Chongqing 401331, China}

\author{Yuchen Gu}
\affiliation{International Center for Quantum Materials, School of Physics, Peking University, Beijing 100871, China}

\author{Aifeng Wang}
\email{afwang@cqu.edu.cn}
\affiliation{College of Physics \& Center of Quantum Materials and Devices, Chongqing University, Chongqing 401331, China}

\author{Yuan Li}
\email{yuan.li@pku.edu.cn}
\affiliation{International Center for Quantum Materials, School of Physics, Peking University, Beijing 100871, China}

\author{Yisheng Chai}
\email{yschai@cqu.edu.cn}
\affiliation{College of Physics \& Center of Quantum Materials and Devices, Chongqing University, Chongqing 401331, China}

\author{Mingquan He}
\email{mingquan.he@cqu.edu.cn}
\affiliation{College of Physics \& Center of Quantum Materials and Devices, Chongqing University, Chongqing 401331, China}

\date{\today}

\begin{abstract}
 In correlated quantum materials, divergent critical fluctuations near the quantum critical point are often closely associated with exotic quantum phases of matter, such as unconventional superconductivity and quantum spin liquids. Here we present a simple yet highly sensitive composite magnetoelectric (ME) method for detecting the critical spin fluctuations in quantum magnets. The ME signal is proportional the magnetostriction coefficient, which directly probes the product of magnetization and spin-spin correlation. As a demonstration, the composite ME method is applied to a Kitaev quantum spin liquid candidate \ce{Na3Co2SbO6}, which shows signs of magnetic field-induced quantum criticality.  Notably, the ME signal prominently diverges at the magnetic field-induced tricritical points, particularly at a tricritical point that lies in close proximity to a zero-temperature quantum critical point. A crucial aspect of these tricritical points is their tunability through the modification of the in-plane magnetic field's direction. The direction of magnetic field can thus serve as a handful yet important tuning parameter, alongside pressure and chemical doping, for searching quantum critical points in quantum magnets with pronounced magnetic anisotropy.

\end{abstract}

\maketitle

\begin{figure}
\includegraphics[width=230pt]{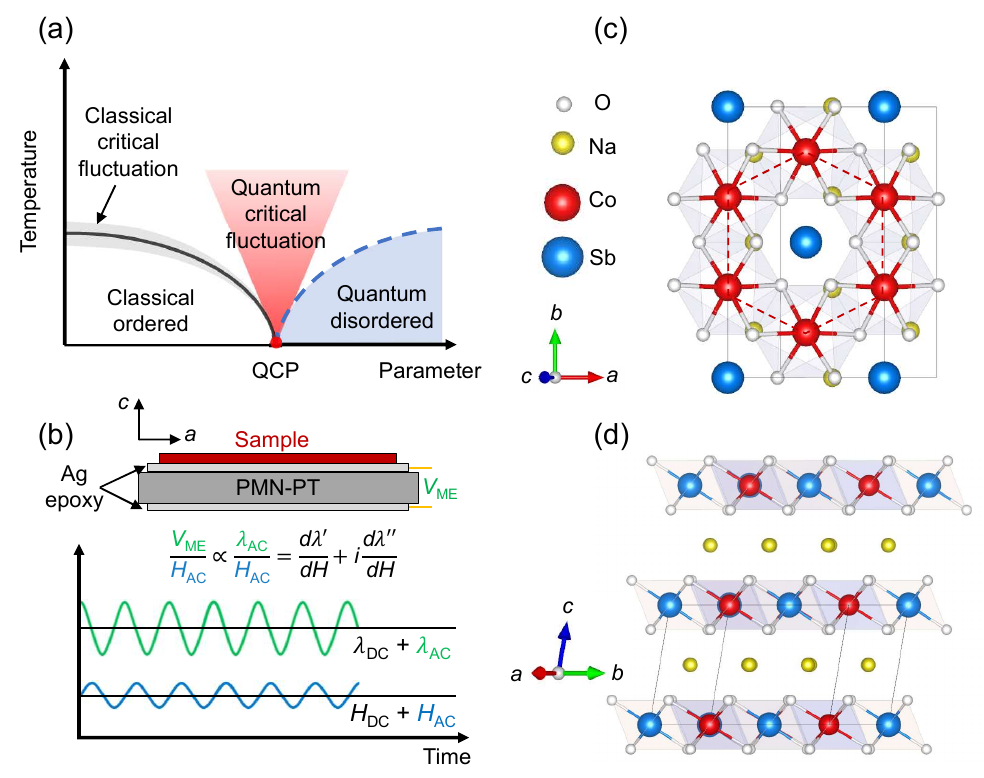}
\caption{\label{fig:1}(a) A schematic phase diagram showing both classical and quantum critical fluctuations. (b) A schematic setup of the composite ME method used to measure the magnetostriction coefficient. A sample is mechanically bonded to a PMN-PT single crystal using Ag epoxy. In the presence of a magnetic field, the magnetostriction ($\lambda$) of the sample is converted to the voltage ($V_\mathrm{ME}$) across the PMN-PT via the piezoelectric effect. By superimposing a small AC component ($H_\mathrm{AC}$) onto the DC magnetic field ($H_\mathrm{DC}$), the AC magnetostriction coefficient $d\lambda_\mathrm{AC}/dH$ of the sample can be probed by monitoring the AC voltage of the PMN-PT: $V_\mathrm{ME}/H_\mathrm{AC}\propto \lambda_\mathrm{AC}/H_\mathrm{AC}=d\lambda'/dH+id\lambda''/dH$ with $\lambda'$ and $\lambda''$ being the real and imaginary parts of $d\lambda_\mathrm{AC}/dH$. (c) Top view and (d) side view of the crystal structure of \ce{Na3Co2SbO6}. Edge sharing \ce{CoO6} octahedra form a honeycomb lattice within the $ab$ plane. } 
\end{figure}

Critical phenomena appear ubiquitously near continuous phase transitions and critical points [see Fig. \ref{fig:1}(a)] \cite{stanley1971phase,Wilson1975}. Coexistence of  indistinguishable phases in the critical region is naturally associated with strong critical fluctuations \cite{Hohenberg1977,Chaikin_Lubensky_1995}. A classic manifestation of such phenomena is critical opalescence, observed at liquid-gas or binary liquid-liquid critical points, where large density fluctuations blur the boundary between different phases \cite{Cagniard1822,Einstein1910}.  Similarly, in quantum spin systems, critical spin fluctuations near quantum critical points (QCPs) can act as a driving force for the emergence of various quantum phases of matter, including unconventional superconductivity, non-Fermi liquid states, or quantum spin liquids \cite{Brando2016,coleman2005quantum,sachdev2000,Senthil2004}.The exploration, identification, and manipulation of these critical spin fluctuations hold profound implications for both our fundamental understanding of quantum materials and potential technological applications \cite{burch2018magnetism,jin2020imaging}. 

As one approaches a magnetic critical point, spin fluctuations can manifest across all spatial and temporal scales, leading to singularities or divergences in several physical quantities such as magnetic susceptibility, specific heat, correlation length, and spin-spin correlation. While experimental measurements of magnetic susceptibility and specific heat are relatively straightforward, employing either commercial equipment or custom setups, the determination of correlation length and spin-spin correlation typically requires access to neutron scattering facilities, which are less readily available. Moreover, neutron scattering experiments are limited in their ability to map out spin-spin correlation across a restricted range of wave vectors and frequencies.  Notably, the strength of magnetic exchange interactions is highly sensitive to the spatial separation between adjacent magnetic moments. This dependency means that the spin-spin correlation $\langle \mathbf{S}_i\cdot\mathbf{S}_j\rangle$ (with $\mathbf{S}_{i,j}$  denoting spin operators at sites ($i,j$) can be directly accessed by magnetostriction $\lambda(H)$, \textit{i.e.}, the change in sample length induced by an external magnetic field (H) \cite{Kittel1960,Callen1965,Callen1968,zapf2008}. The magnetostriction is a thermodynamic quantity that can be measured with high precision using various homemade setups, including capacitance dilatometer \cite{kuchler2012}, optical fiber Bragg grating \cite{Daou2010} and  piezoresistive cantilever \cite{Park2009},  thereby offering a viable method for studying spin fluctuations. This approach to accessing spin-spin correlations via magnetostriction has been successfully applied in a variety of magnetic systems. Notable examples include spin-chain magnet \ce{NiCl2-4SC(NH2)2} \cite{zapf2008}, quasi-one-dimensional spin-1/2 magnet \ce{LiCuVO4} \cite{Ikeda2019} and kagome spin-1/2 magnet \ce{Cu3V2O7(OH)2}$\cdot$\ce{2H2O} \cite{Miyata2021}.

Even more interesting with this respect is the magnetostriction coefficient $d\lambda(H)/dH$, which is proportional to the product of magnetization and spin-spin correlation \cite{Callen1965,Callen1968}: 

\begin{equation}
    \frac{d\lambda(H)}{dH} \sim \sum\limits_{k\langle i,j\rangle}D(i,j)\langle S_k^z \mathbf{S}_i\cdot\mathbf{S}_j\rangle,
\label{eq:1}
\end{equation}
where $D(i,j)$ is the magnetoelastic constant between sites ($i,j$), $\langle S_k^z \rangle$ is the average magnetization with the external magnetic field applied along the $z$ direction. In the critical region, the divergence of both $\langle S_k^z \rangle$  and $\langle \mathbf{S}_i\cdot\mathbf{S}_j\rangle$  renders the magnetostriction coefficient exceptionally responsive to spin fluctuations.

\begin{figure}
\includegraphics[scale=0.35]{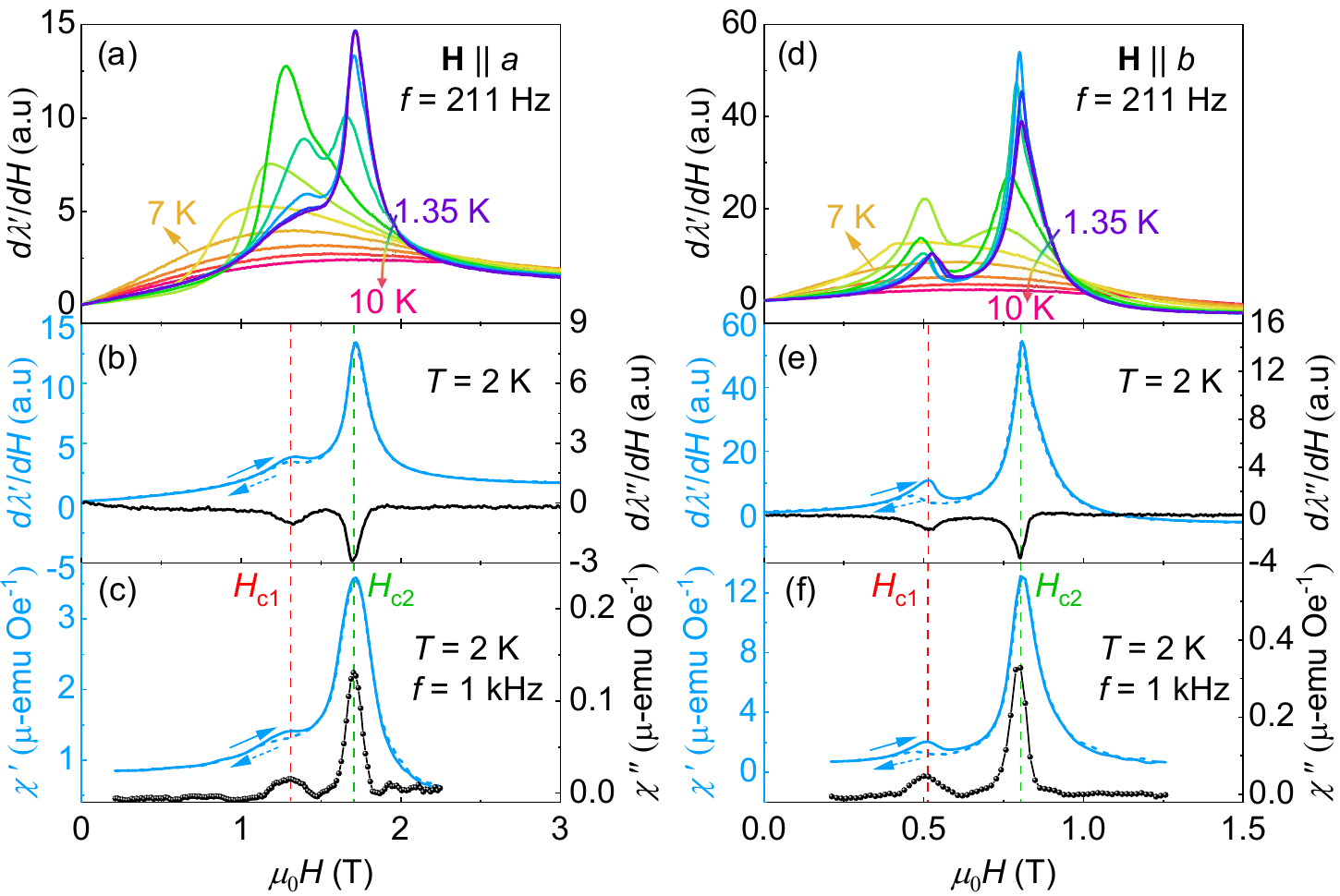}
\caption{\label{fig:3} (a) The magnetostriction coefficient measured as a function of magnetic field captured at selected temperatures ($\mathbf{H}\parallel a$). (b) and (c) Comparison of the real and imaginary parts of the ac magnetostriction coefficient and ac magnetic susceptibility recorded at 2 K ($\mathbf{H}\parallel a$). (d-f) Similar to (a-c) but with $\mathbf{H}\parallel b$. Solid and dashed arrows in (b,c) and (e,f) represent increasing and decreasing magnetic fields, respectively.   }
\end{figure}

In this context, we introduce a straightforward yet sensitive technique for measuring the magnetostriction coefficient and thus probing spin fluctuations in quantum magnets. As shown in Fig. \ref{fig:1}(b), this method is based on the composite magnetoelectric (ME) effect \cite{Chai2021,zhang2023}. The method involves mechanically coupling the sample to a piezoelectric single crystal, 0.7Pb(\ce{Mg1/3Nb2/3})\ce{O3} - 0.3\ce{PbTiO3} (PMN-PT), using silver epoxy. The PMN-PT crystals were shaped into [001]-cut thin plates of 0.2 mm thickness and electrically poled at room temperature with a 550 kV/m electric field.  An applied external magnetic field induces magnetostrictive deformation in the sample, which is transferred to the PMN-PT, generating an electrical voltage  $V_\mathrm{ME}$ via the piezoelectric effect. Here, PMN-PT serves as a strain sensor, making $V_\mathrm{ME}$  a direct indicator of the magnetostriction coefficient:

\begin{equation}
    V_\mathrm{ME} \propto k\frac{dE}{d\lambda}\frac{d\lambda}{dH},
\label{eq:2}
\end{equation}
with $0<k<1$ reflecting the efficiency of strain transfer between the sample and PMN-PT, $dE/d\lambda$ being the piezoelectric coefficient of PMN-PT. To enhance sensitivity, an AC magnetic field (1 Oe) generated by a custom Helmholtz coil is superimposed onto the DC magnetic field. The resulting voltage, $V_\mathrm{ME}$,  is an AC signal that can be precisely measured using lock-in techniques. This approach allows for the simultaneous measurement of both the real ($d\lambda'/dH$)  and imaginary ($d\lambda''/dH$) parts of the magnetostriction coefficient. Specifically, $V_\mathrm{ME}/H_\mathrm{AC}\propto \lambda_\mathrm{AC}/H_\mathrm{AC}=d\lambda'/dH+id\lambda''/dH$. This AC ME method is extremely useful for identifying critical points that connect second-order and first-order phase transitions. First-order phase transitions are often associated with dissipation, which can be effectively indicated by the imaginary component of the magnetostriction coefficient, similar to AC susceptibility \cite{Zhang_disspation,zhang2023,Topping_AC} [see Fig. \ref{fig:3} for example].

\begin{figure}
\includegraphics[scale=0.35]{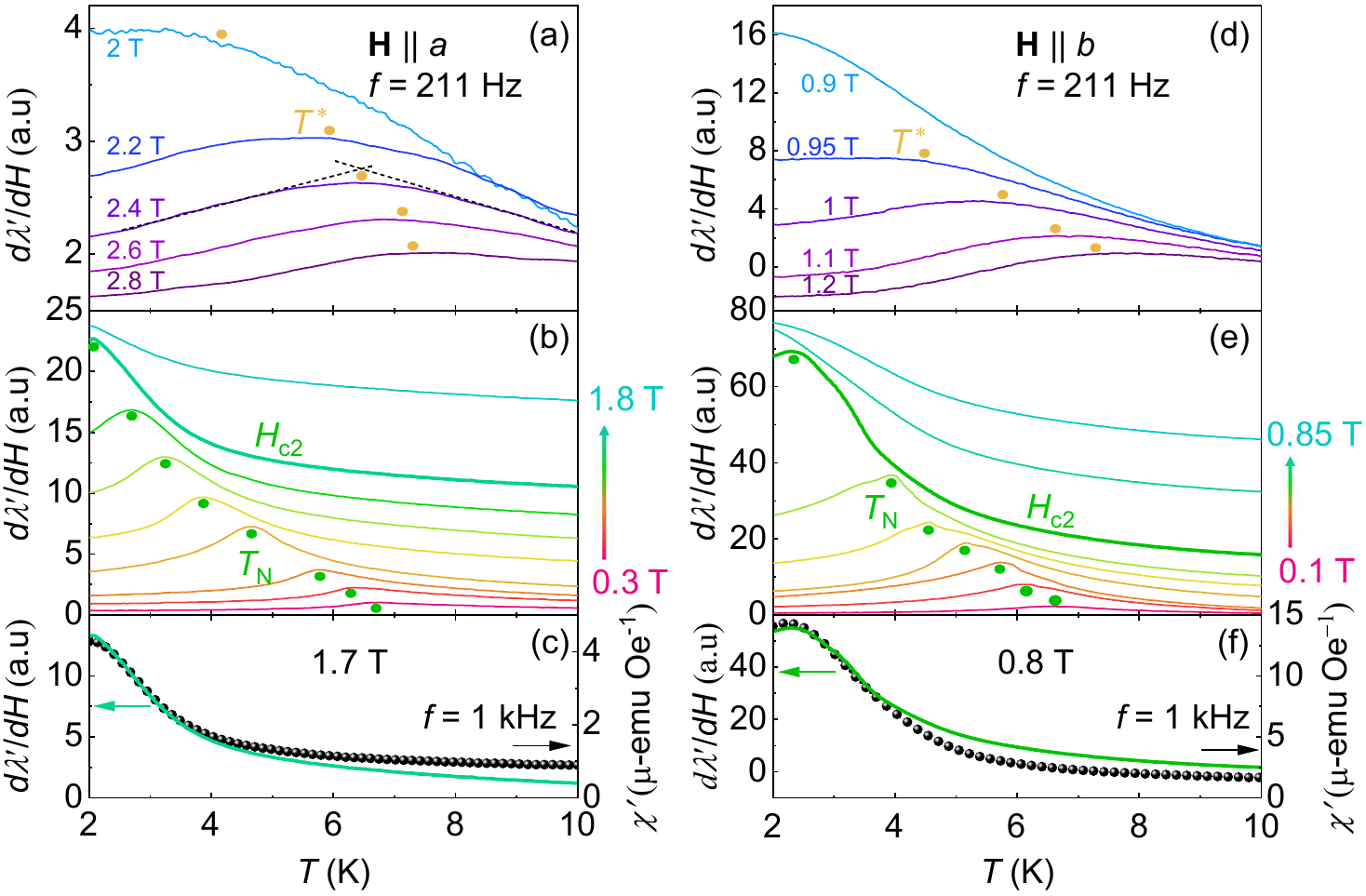}
\caption{\label{fig:2} (a) and (b) Temperature-dependent  magnetostriction coefficient measured with $\mathbf{H}\parallel a$.  (c) Comparison of the real parts of the AC magnetostriction coefficient ($d\lambda'/dH$, green lines) and the AC magnetic susceptibility ($\chi'$, black dots) measured at $H_\mathrm{c2}$. (d-f) Similar to (a-c) but with $\mathbf{H}\parallel b$. Curves in (b) and (e) have been shifted vertically for clarity. The green dots in (b) and (e) mark the AFM transition temperature $T_\mathrm{N}$. The crossover temperature  $T^{*}$ between the field-polarized FM and PM states is labeled by orange dots, chosen as the intersecting points of the linear extrapolated lines of the high- and low-temperature data [see dashed lines in (a)]. }
\end{figure}

To illustrate the effectiveness of the magnetostriction coefficient in detecting critical spin fluctuations, we applied the composite ME method to a Kitaev quantum spin liquid (QSL) candidate, \ce{Na3Co2SbO6}. This approach revealed a clearly divergent ME signal at magnetic field-induced tricritical points, underscoring the composite ME method's potential as a prominent technique for identifying critical fluctuations in quantum magnets.

The Kitaev QSL represents an exactly solvable model characterized by bond-dependent spin-1/2 Ising interactions on a two-dimensional honeycomb lattice \cite{KITAEV20062}. Notably, the Kitaev QSL features anyonic excitations, which hold promise for applications in topological quantum computing \cite{KITAEV20032}. The quest for material realizations of the Kitaev QSL has led to the identification of several spin-orbital assisted Mott insulators, including \ce{(Li, Na)2IrO3}\cite{PhysRevLett.108.127203,PhysRevLett.110.097204,hwan2015direct,PhysRevB.82.064412}, $\alpha-$\ce{RuCl3}\cite{banerjee2016proximate,Banerjee2017science,PhysRevB.90.041112,kasahara2018majorana}, \ce{BaCo2(AsO4)2}\cite{PhysRevB.104.144408,ruidansciadv.aay6953,zhang2023magnetic}, \ce{Na2Co2TeO6} and \ce{Na3Co2SbO6}\cite{VICIU20071060,Li.L180404,lin2021field,Songvilay224429,Yan074405}. However, real materials often host additional interactions beyond the Kitaev model, culminating in long-range magnetic order. Despite this, the suppression of long-range order via external parameters like magnetic fields has unveiled signatures of Kitaev QSL behaviors in these materials. Among such candidates, \ce{Na3Co2SbO6} emerges as a particularly promising example, distinguished by its relatively low magnetic ordering temperature and the modest magnetic field required to mitigate magnetic order \cite{Li2022,Vavi2023,Hu2024}. Furthermore, signatures of quantum criticality have been found in \ce{Na3Co2SbO6} when its magnetic order is suppressed by magnetic field \cite{Hu2024}, thus making \ce{Na3Co2SbO6} a prime subject for studying critical spin fluctuations using the composite ME method.   

As shown in Figs. \ref{fig:1}(c,d), \ce{Na3Co2SbO6} features a monoclinic layered structure (space group $C2/m$) \cite{VICIU20071060}. Co atoms form a honeycomb lattice within each layer along  the $ab$-plane. Adjacent \ce{[Co2SbO6]}$^{3-}$ layers are stacked along the $c$-axis, separated by \ce{Na+} ions. The high-spin state of the Co$^{2+}$ (3$d^7$)  leads to a pseudospin  $j_\mathrm{eff}=1/2$, facilitated by spin-orbit couplings. Dominant Kitaev interactions are theoretically expected due to the involvement of $t_{2g}-e_g$ and $e_g-e_g$ exchange paths \cite{Liu2020}, as also evidenced by neutron scattering measurements \cite{Songvi2020,Sanders2022,Kim_2022}.  Like other Kitaev materials, \ce{Na3Co2SbO6} enters a long-range antiferromagnetic (AFM) state below $T_\mathrm{N}\sim 5-8$ K depending on sample quality \cite{VICIU20071060,WONG201618,Yan2019,C9NJ03627J,Li2022,Hu2024,Vavi2023,veenendaal2023}. 

Despite the minor orthorhombic distortion ($<0.2$\%) within the $ab$ plane, giant in-plane magnetic anisotropy has been found by magnetization and neutron diffraction experiments \cite{Li2022}. To resolve the in-plane anisotropy using the composite ME method, a twin-free single crystal was employed. Figure \ref{fig:2} displays the temperature-dependent ME signal with external magnetic fields applied along different in-plane directions. In small magnetic fields, $d\lambda'/dH$ increases rapidly upon cooling and drops sharply at $T_\mathrm{N}$, akin to magnetic susceptibility. In 0.1 T, the transition temperature is found to be $T_\mathrm{N}=6.7$ K, agreeing well with our earlier study \cite{Li2022}.  In both configurations, $T_\mathrm{N}$ shifts gradually towards lower temperatures with increasing magnetic fields. At a critical field of $\mu_0H_\mathrm{c2}\sim$ 1.8 T (0.8 T) in the $\mathbf{H} \parallel a$ ($\mathbf{H} \parallel b$) configuration, the AFM transition is suppressed below 2 K. The large differences in $H_\mathrm{c2}$ for different in-plane field directions indicate large in-plane magnetic anisotropy, as also revealed by magnetization and neutron scattering experiments \cite{Li2022}. As shown in Figs. \ref{fig:2}(c,f), near the critical field $H_\mathrm{c2}$ where spin fluctuations are most profound, the magnetostriction coefficient and the AC magnetic susceptibility  scale nicely with each other, further demonstrating that the magnetostriction coefficient directly probes spin fluctuations. In higher magnetic fields well above $H_\mathrm{c2}$, the long-range AFM state is suppressed and a field-polarized ferromagnetic (FM) state is reached at low temperatures. In the high-field region, a broad hump peaking at $T^*$ appears in the temperature-dependent $d\lambda'/dH$ signal, which signifies the crossover between the paramagnetic (PM) state and the field-polarized FM phase. Similar crossover behavior near $T^*$ is also observed in AC magnetic susceptibility [see Fig. S1 Supplemental Material \cite{supplemental}]. 

\begin{figure*}
\includegraphics[width=500pt]{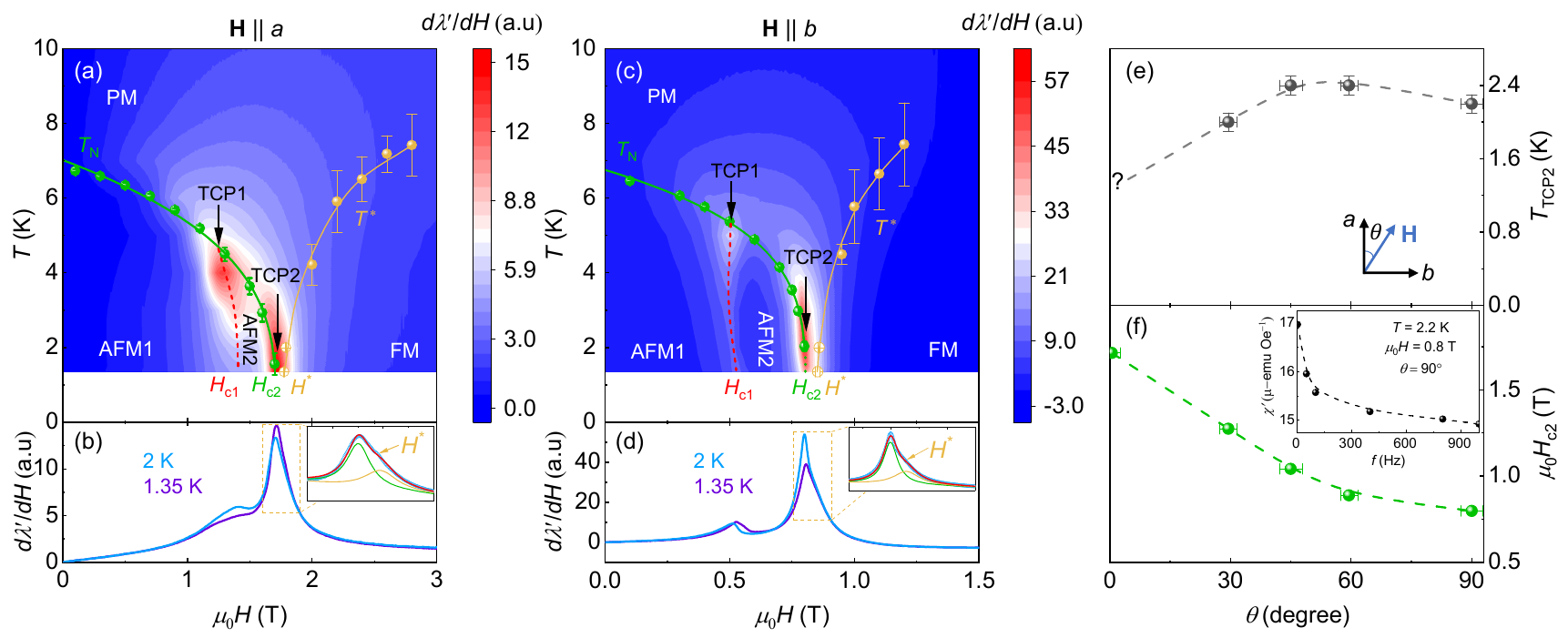}
\caption{\label{fig:4} (a) and (c) Phase diagrams mapped using the magnetostriction data for $\mathbf{H}\parallel a$ and $\mathbf{H}\parallel b$. Dashed and solid lines represent first- and second-order phase transitions, respectively. (b) and (d) Comparison of $d\lambda'/dH$ measured at 2 and 1.35 K. Insets in (b) and (d) show magnified view near $H_{c2}$ and $H^*$. The red solid lines in the insets of (b) and (d) fit the $d\lambda'/dH$ (2 K) data using Lorentzian profiles for the transitions at $H_\mathrm{c2}$ (green lines) and $H^{*}$ (orange lines). (e) and (f) Angular dependence of the critical temperature $T_\mathrm{TCP2}$ and critical field $H_\mathrm{c2}$ of TCP2 obtained from AC magnetic susceptibility data. The inset in (f) shows the frequency dependency of $\chi'$ at TCP2 with $\mathbf{H}\parallel b$.}
\end{figure*}

To further study the magnetic field-induced transitions, we present the magnetic field-dependent $d\lambda'/dH$ data in Fig. \ref{fig:3}. Well above the AFM transition, $d\lambda'/dH$ only varies weakly with magnetic field. 
As $T_\mathrm{N}$ is approached with cooling, the development of spin fluctuations leads to a broad hump in $d\lambda'/dH$ between approximately 1 and 2 T (0.5 and 1 T) for $\mathbf{H} \parallel a$ ($\mathbf{H} \parallel b$).  Below $T_\mathrm{N}$, two distinct peaks emerge at $H_\mathrm{c1}$ and $H_\mathrm{c2}$, indicative of field-induced transitions.  The $d\lambda'/dH$ signal is most intense near  $T_\mathrm{N}$ for the transition at $H_\mathrm{c1}$, while for the transition at $H_\mathrm{c2}$, $d\lambda'/dH$ significantly increases at lower temperatures. These transitions have also been clearly observed in AC magnetic susceptibility, as shown in Figs. \ref{fig:3}(c) and \ref{fig:3}(f). Notably, a clear hysteresis near  $H_\mathrm{c1}$ is observed both in $d\lambda'/dH$ and $\chi'$ when comparing  the data measured with increasing and decreasing magnetic fields [see Figs. \ref{fig:3}(b,c) and \ref{fig:3}(e,f)].  A similar hysteresis behavior is also present in DC magnetization measurements, suggesting a first-order transition at $H_\mathrm{c1}$ \cite{Li2022}. Moreover, a peak feature is also seen in the imaginary part both in $d\lambda''/dH$ and $\chi''$, suggesting finite dissipation and further pointing to the first-order nature of the transition at $H_\mathrm{c1}$. Above $H_\mathrm{c1}$, \ce{Na3Co2SbO6} transitions into an AFM state characterized by different wave vectors than those in zero field, as evidenced by neutron diffraction experiments \cite{Li2022}. Here, we denote the phases below and above $H_\mathrm{c1}$ as AFM1 and AFM2, respectively. In higher magnetic fields well above $H_\mathrm{c2}$, the system ultimately reaches a field-polarized FM state.  Unlike the transition at $H_\mathrm{c1}$, no hysteresis is observed at $H_\mathrm{c2}$ down to 1.35 K for both field configurations. However, clear peaks are found at $H_\mathrm{c2}$ in  $d\lambda''/dH$ and $\chi''$ at low temperatures, indicative to sizable dissipation likely originated from a first-order transition [see Figs. \ref{fig:3}(b,c) and \ref{fig:3}(e,f)]. Indeed, near $H_\mathrm{c2}$, neutron scatting measurements performed with $\mathrm{B}\parallel a$ have clearly found coexistence of the AFM2 and the field-polarized FM state at 0.25 K \cite{Li2022}. A recent nuclear magnetic resonance (NMR) study also hints at a weak first-order quantum phase transition at $H_\mathrm{c2}$ \cite{Hu2024}. The absence of hysteresis in a first-order transition is not uncommon, especially when the energy barrier separating two phases is smaller than the energy associated with fluctuations \cite{Guillou_first,Alho_first,Biswas_first}. During the process of a first-order transition, migration of phase interfaces between coexisted phases act as internal friction, which causes dissipation \cite{Zhang_disspation,zhang2023,Topping_AC}. This dynamic process can be nicely captured by AC probes, such as AC magnetic susceptibility and our AC magnetostriction method presented here. While a second-order phase transition could produce an imaginary response due to a resonating soft mode with the AC probe, such phenomena typically occur at much higher frequencies than those used in our measurements and are unlikely to explain our observed data.  Note that in the high-temperature region above 3 K (4 K) with $\mathbf{H} \parallel a$ ($\mathbf{H} \parallel b$), the imaginary part $d\lambda''/dH$ is featureless at $H_\mathrm{c2}$, suggestive to a second-order transition [see Figs.S3 and S4 in Supplemental Material \cite{supplemental}]. These findings thus imply the existence of a tricritical point that connects the high-temperature second-order transition line with the low-temperature first-order line at $H_\mathrm{c2}$.

In Fig. \ref{fig:4}, we summarize the phase diagrams derived from the composite ME method.  Similar phase diagrams are obtained using AC magnetic susceptibility measurements [see Fig. S2 in Supplemental Material \cite{supplemental}]. For both  $\mathbf{H} \parallel a$ and $\mathbf{H} \parallel b$ configurations, $T_\mathrm{N}$ is suppressed continuously with increasing magnetic fields. The field dependence of $T_\mathrm{N}(H)$ can be nicely described by a power law scaling $T_\mathrm{N}(H)\sim(1-H/H_\mathrm{c2})^{z\nu}$, as depicted by the solid green lines in  Figs. \ref{fig:4}(a) and \ref{fig:4}(c). The scaling gives a critical field of $H_\mathrm{c2}=1.71\pm0.01$ ($H_\mathrm{c2}=0.84\pm0.01$ ) and a critical exponent $z\nu=0.32\pm0.01$ ($z\nu=0.24\pm0.01$) for  $\mathrm{H}\parallel a$ ($\mathrm{H}\parallel b$). The critical exponent is consistent with that reported in polycrystalline samples using magnetization experiments \cite{Vavi2023}.  More notably, the $d\lambda'/dH$ signal intensifies significantly near the transitions at  $H_\mathrm{c1}$ and $H_\mathrm{c2}$, with two distinct maxima observable in each phase diagram displayed as the contour plot of the data shown in Fig. \ref{fig:3}. The first maximum in $d\lambda'/dH$ emerges at the juncture of the second-order $T_\mathrm{N}(H)$ line and the first-order $H_\mathrm{c1}(T)$ line. This convergence point is thus a tricritical point (TCP), denoted as TCP1 in Figs. \ref{fig:4}(a,b).  In addition, three phases including PM, AFM1 and AFM2 converge at this point. The coexistence of multiple phases triggers intense fluctuations, leading to divergent magnetic susceptibility and  magnetostriction coefficient.Note that the slight deviation of the diverging point from TCP1 is caused by the coarse temperature step (1 K) used in the magnetic field scan presented in Fig. \ref{fig:3}. It can be observed that the diverging point is located exactly at TCP1 when finer temperature steps are used, as evidenced by AC susceptibility data shown in Fig. S2 Supplemental Material \cite{supplemental}.   

Similarly, another TCP (labeled as TCP2) emerges at the second maximum in $d\lambda'/dH$, which connects the high-temperature second order and low-temperature first-order $T_\mathrm{N}(H)$ lines. Compared to TCP1, the intensity of the $V_\mathrm{ME}$ signal near TCP2 is significantly stronger, especially for the configuration with $\mathbf{H}\parallel b$. This enhancement can be attributed to strong quantum fluctuations in the vicinity of TCP2, as evidenced by nuclear NMR experiments that detected signs of QSL behaviors and quantum criticality \cite{Hu2024,Vavi2023}. Hence, TCP2 resides within a quantum critical region, which leads to an amplified $V_\mathrm{ME}$ signal in comparison to the classical criticality observed at TCP1. With both of the in-plane field geometry, we believe that TCP2 is situated at finite temperatures, hence giving rise to the first-order characteristics we observed in the dissipative (imaginary part of) ME signals. An intriguing direction for future research would be to explore methods to lower TCP2 to zero temperature, thereby potentially realizing a quantum tricritical point (QTCP).  Achieving a QTCP, especially given the first-order nature of the transition between the two ordered states (AFM2 and FM), may also facilitate the emergence of a deconfined QTCP. It is important to note the sensitivity of TCP2 to the direction of the applied in-plane magnetic field, a consequence of the significant in-plane magnetic anisotropy in \ce{Na3Co2SbO6}.  As shown in Figs. \ref{fig:4}(e,f), both the critical temperature ($T_\mathrm{TCP2}$) and critical magnetic field ($H_\mathrm{c2}$) of TCP2 vary significantly with the direction of the in-plane magnetic field. A similar angular dependency is also found in the phonon thermal conductivity ($\kappa_\mathrm{ph}$) of \ce{Na3Co2SbO6}, in which $\kappa_\mathrm{ph}$ measured at 0.3 K is largest when the magnetic field  ($H=H_\mathrm{c2}$) is directed 50 degrees away from the $a$-axis ($\theta=50^{\circ}$) \cite{Fan_kappa}. As seen in Fig. \ref{fig:4}(e), for $\theta=50^{\circ}$, $T_\mathrm{TCP2}$ is furthest from 0 K, and spin fluctuations are expected to be the weakest at 0.3 K compared to other directions. This naturally explains the largest $\kappa_\mathrm{ph}$ due to the weakest spin-phonon scattering at 0.3 K for $\theta=50^{\circ}$.  Note that for $\mathbf{H}\parallel a$, $T_\mathrm{TCP2}$ can not be determined in the present study. As shown in Fig. \ref{fig:4}(b), at $H_\mathrm{c2}$, $d\lambda'/dH$ continues to increase with cooling from 2 K to 1.35 K. In addition, a shoulder feature is found slightly above $H_\mathrm{c2}$ near $H^{*}$, indicating the crossover between PM and field-polarized FM phases, or the ultimate disappearance of the AFM2 state.  Previous neutron study has shown that the AFM2 and FM coexists at 0.25 K near $H_\mathrm{c2}$ for $\mathbf{H}\parallel a$ \cite{Li2022}. This suggests that for $\mathbf{H}\parallel a$, the $T_\mathrm{N}(H)$ and $T^{*}(H)$ lines have crossed (if not merged together) somewhere between 1.35 and 0.25 K, with TCP2 likely located at the crossing point. Although a QTCP is not realized  here in \ce{Na3Co2SbO6}, the findings provide a crucial insight into manipulating critical points and pursuing QCP by varying the direction of the magnetic field in quantum magnets with pronounced anisotropy. Similar tunable critical points by the direction of the magnetic field have been found in the itinerant metamagnet \ce{Sr3Ru2O7}, which shows non-Fermi liquid behavior near the field-induced QCP with $\mathbf{H}\parallel c$ \cite{SrRuO,Sr3Ru2O7_angle}. Compared to conventional tuning parameters like pressure and chemical doping, adjusting the magnetic field direction is considerably more straightforward. Utilizing the composite ME method alongside this approach enables the efficient exploration and characterization of various critical points, including both classical and quantum critical points.

A final remark about the AC magnetostriction coefficient and AC magnetic susceptibility measurements is that although they are extremely sensitive to critical spin fluctuations, unveiling the nature of a critical point remains challenging. As displayed in the inset of Fig. \ref{fig:4}(f), at TCP2, the intensity of  $\chi'$ is highly frequency-dependent and diverges sharply in the zero-frequency limit. Although a relatively low frequency (211 Hz) is used in the AC magnetostriction coefficient experiments, it is still far from the static limit. The critical slowing down at finite frequencies can originate from disorder and/or the presence of domain walls near the first-order transition \cite{Sr3Ru2O7_angle,disorder}. Consequently, extracting the critical exponents with great certainty from the temperature- and field-dependent AC magnetostriction coefficient and AC magnetic susceptibility measured at finite frequency is difficult. 
 
In summary, we have introduced a straightforward yet highly sensitive technique for probing critical spin fluctuations through the measurement of the magnetostriction coefficient via a composite magnetoelectric (ME) method. Utilizing this approach, we have successfully delineated the intricate magnetic phase diagrams of the Kitaev material \ce{Na3Co2SbO6}. Notably, we identified two tricritical points within these diagrams, where the ME signal sharply diverges under the influence of in-plane magnetic fields. Interestingly, one of these tricritical points is in close proximity to a quantum critical point and demonstrates considerable tunability through adjustments in the direction of the in-plane magnetic field. This offers an alternative avenue for exploring quantum phase transitions, complementing traditional methods such as pressure and doping, particularly in quantum magnets characterized by significant magnetic anisotropy. The composite ME method thereby serves as a powerful tool for advancing the investigation of quantum critical points as well as the associated quantum critical spin fluctuations.

\medskip

\section*{Acknowledgements}

We thank Dietrich Belitz for fruitful discussions. This work has been supported by National Natural Science Foundation of China (Grants No. 12104254),  Chinesisch-Deutsche Mobilit\"atsprogamm of Chinesisch-Deutsche Zentrum f\"ur Wissenschaftsf\"orderung (Grant No. M-0496), the Open Fund of the China Spallation Neutron Source Songshan Lake Science City. The work at Peking University has been supported in part by the National Basic Research Program of China (Grants No. 2021YFA1401900).

\nocite{*}


\begin{thebibliography}{62}%
\makeatletter
\providecommand \@ifxundefined [1]{%
 \@ifx{#1\undefined}
}%
\providecommand \@ifnum [1]{%
 \ifnum #1\expandafter \@firstoftwo
 \else \expandafter \@secondoftwo
 \fi
}%
\providecommand \@ifx [1]{%
 \ifx #1\expandafter \@firstoftwo
 \else \expandafter \@secondoftwo
 \fi
}%
\providecommand \natexlab [1]{#1}%
\providecommand \enquote  [1]{``#1''}%
\providecommand \bibnamefont  [1]{#1}%
\providecommand \bibfnamefont [1]{#1}%
\providecommand \citenamefont [1]{#1}%
\providecommand \href@noop [0]{\@secondoftwo}%
\providecommand \href [0]{\begingroup \@sanitize@url \@href}%
\providecommand \@href[1]{\@@startlink{#1}\@@href}%
\providecommand \@@href[1]{\endgroup#1\@@endlink}%
\providecommand \@sanitize@url [0]{\catcode `\\12\catcode `\$12\catcode `\&12\catcode `\#12\catcode `\^12\catcode `\_12\catcode `\%12\relax}%
\providecommand \@@startlink[1]{}%
\providecommand \@@endlink[0]{}%
\providecommand \url  [0]{\begingroup\@sanitize@url \@url }%
\providecommand \@url [1]{\endgroup\@href {#1}{\urlprefix }}%
\providecommand \urlprefix  [0]{URL }%
\providecommand \Eprint [0]{\href }%
\providecommand \doibase [0]{https://doi.org/}%
\providecommand \selectlanguage [0]{\@gobble}%
\providecommand \bibinfo  [0]{\@secondoftwo}%
\providecommand \bibfield  [0]{\@secondoftwo}%
\providecommand \translation [1]{[#1]}%
\providecommand \BibitemOpen [0]{}%
\providecommand \bibitemStop [0]{}%
\providecommand \bibitemNoStop [0]{.\EOS\space}%
\providecommand \EOS [0]{\spacefactor3000\relax}%
\providecommand \BibitemShut  [1]{\csname bibitem#1\endcsname}%
\let\auto@bib@innerbib\@empty
\bibitem [{\citenamefont {Stanley}(1971)}]{stanley1971phase}%
  \BibitemOpen
  \bibfield  {author} {\bibinfo {author} {\bibfnamefont {H.~E.}\ \bibnamefont {Stanley}},\ }\href@noop {} {\emph {\bibinfo {title} {Phase transitions and critical phenomena}}},\ Vol.~\bibinfo {volume} {7}\ (\bibinfo  {publisher} {Clarendon Press, Oxford},\ \bibinfo {year} {1971})\BibitemShut {NoStop}%
\bibitem [{\citenamefont {Wilson}(1975)}]{Wilson1975}%
  \BibitemOpen
  \bibfield  {author} {\bibinfo {author} {\bibfnamefont {K.~G.}\ \bibnamefont {Wilson}},\ }\href {https://doi.org/10.1103/RevModPhys.47.773} {\bibfield  {journal} {\bibinfo  {journal} {Rev. Mod. Phys.}\ }\textbf {\bibinfo {volume} {47}},\ \bibinfo {pages} {773} (\bibinfo {year} {1975})}\BibitemShut {NoStop}%
\bibitem [{\citenamefont {Hohenberg}\ and\ \citenamefont {Halperin}(1977)}]{Hohenberg1977}%
  \BibitemOpen
  \bibfield  {author} {\bibinfo {author} {\bibfnamefont {P.~C.}\ \bibnamefont {Hohenberg}}\ and\ \bibinfo {author} {\bibfnamefont {B.~I.}\ \bibnamefont {Halperin}},\ }\href {https://doi.org/10.1103/RevModPhys.49.435} {\bibfield  {journal} {\bibinfo  {journal} {Rev. Mod. Phys.}\ }\textbf {\bibinfo {volume} {49}},\ \bibinfo {pages} {435} (\bibinfo {year} {1977})}\BibitemShut {NoStop}%
\bibitem [{\citenamefont {Chaikin}\ and\ \citenamefont {Lubensky}(1995)}]{Chaikin_Lubensky_1995}%
  \BibitemOpen
  \bibfield  {author} {\bibinfo {author} {\bibfnamefont {P.~M.}\ \bibnamefont {Chaikin}}\ and\ \bibinfo {author} {\bibfnamefont {T.~C.}\ \bibnamefont {Lubensky}},\ }\href@noop {} {\emph {\bibinfo {title} {Principles of Condensed Matter Physics}}}\ (\bibinfo  {publisher} {Cambridge University Press},\ \bibinfo {year} {1995})\BibitemShut {NoStop}%
\bibitem [{\citenamefont {Cagniard de~la Tour}(1822)}]{Cagniard1822}%
  \BibitemOpen
  \bibfield  {author} {\bibinfo {author} {\bibfnamefont {C.}~\bibnamefont {Cagniard de~la Tour}},\ }\href@noop {} {\bibfield  {journal} {\bibinfo  {journal} {Ann. Chim Phys.}\ }\textbf {\bibinfo {volume} {21}},\ \bibinfo {pages} {127} (\bibinfo {year} {1822})}\BibitemShut {NoStop}%
\bibitem [{\citenamefont {Einstein}(1910)}]{Einstein1910}%
  \BibitemOpen
  \bibfield  {author} {\bibinfo {author} {\bibfnamefont {A.}~\bibnamefont {Einstein}},\ }\href {https://doi.org/https://doi.org/10.1002/andp.19103381612} {\bibfield  {journal} {\bibinfo  {journal} {Annalen der Physik}\ }\textbf {\bibinfo {volume} {338}},\ \bibinfo {pages} {1275} (\bibinfo {year} {1910})}\BibitemShut {NoStop}%
\bibitem [{\citenamefont {Brando}\ \emph {et~al.}(2016)\citenamefont {Brando}, \citenamefont {Belitz}, \citenamefont {Grosche},\ and\ \citenamefont {Kirkpatrick}}]{Brando2016}%
  \BibitemOpen
  \bibfield  {author} {\bibinfo {author} {\bibfnamefont {M.}~\bibnamefont {Brando}}, \bibinfo {author} {\bibfnamefont {D.}~\bibnamefont {Belitz}}, \bibinfo {author} {\bibfnamefont {F.~M.}\ \bibnamefont {Grosche}},\ and\ \bibinfo {author} {\bibfnamefont {T.~R.}\ \bibnamefont {Kirkpatrick}},\ }\href {https://doi.org/10.1103/RevModPhys.88.025006} {\bibfield  {journal} {\bibinfo  {journal} {Rev. Mod. Phys.}\ }\textbf {\bibinfo {volume} {88}},\ \bibinfo {pages} {025006} (\bibinfo {year} {2016})}\BibitemShut {NoStop}%
\bibitem [{\citenamefont {Coleman}\ and\ \citenamefont {Schofield}(2005)}]{coleman2005quantum}%
  \BibitemOpen
  \bibfield  {author} {\bibinfo {author} {\bibfnamefont {P.}~\bibnamefont {Coleman}}\ and\ \bibinfo {author} {\bibfnamefont {A.~J.}\ \bibnamefont {Schofield}},\ }\href {https://doi.org/10.1038/nature03279} {\bibfield  {journal} {\bibinfo  {journal} {Nature}\ }\textbf {\bibinfo {volume} {433}},\ \bibinfo {pages} {226} (\bibinfo {year} {2005})}\BibitemShut {NoStop}%
\bibitem [{\citenamefont {Sachdev}(2000)}]{sachdev2000}%
  \BibitemOpen
  \bibfield  {author} {\bibinfo {author} {\bibfnamefont {S.}~\bibnamefont {Sachdev}},\ }\href {https://doi.org/10.1126/science.288.5465.475} {\bibfield  {journal} {\bibinfo  {journal} {Science}\ }\textbf {\bibinfo {volume} {288}},\ \bibinfo {pages} {475} (\bibinfo {year} {2000})}\BibitemShut {NoStop}%
\bibitem [{\citenamefont {Senthil}\ \emph {et~al.}(2004)\citenamefont {Senthil}, \citenamefont {Vishwanath}, \citenamefont {Balents}, \citenamefont {Sachdev},\ and\ \citenamefont {Fisher}}]{Senthil2004}%
  \BibitemOpen
  \bibfield  {author} {\bibinfo {author} {\bibfnamefont {T.}~\bibnamefont {Senthil}}, \bibinfo {author} {\bibfnamefont {A.}~\bibnamefont {Vishwanath}}, \bibinfo {author} {\bibfnamefont {L.}~\bibnamefont {Balents}}, \bibinfo {author} {\bibfnamefont {S.}~\bibnamefont {Sachdev}},\ and\ \bibinfo {author} {\bibfnamefont {M.~P.~A.}\ \bibnamefont {Fisher}},\ }\href {https://doi.org/10.1126/science.1091806} {\bibfield  {journal} {\bibinfo  {journal} {Science}\ }\textbf {\bibinfo {volume} {303}},\ \bibinfo {pages} {1490} (\bibinfo {year} {2004})}\BibitemShut {NoStop}%
\bibitem [{\citenamefont {Burch}\ \emph {et~al.}(2018)\citenamefont {Burch}, \citenamefont {Mandrus},\ and\ \citenamefont {Park}}]{burch2018magnetism}%
  \BibitemOpen
  \bibfield  {author} {\bibinfo {author} {\bibfnamefont {K.~S.}\ \bibnamefont {Burch}}, \bibinfo {author} {\bibfnamefont {D.}~\bibnamefont {Mandrus}},\ and\ \bibinfo {author} {\bibfnamefont {J.-G.}\ \bibnamefont {Park}},\ }\href {https://doi.org/10.1038/s41586-018-0631-z} {\bibfield  {journal} {\bibinfo  {journal} {Nature}\ }\textbf {\bibinfo {volume} {563}},\ \bibinfo {pages} {47} (\bibinfo {year} {2018})}\BibitemShut {NoStop}%
\bibitem [{\citenamefont {Jin}\ \emph {et~al.}(2020)\citenamefont {Jin}, \citenamefont {Tao}, \citenamefont {Kang}, \citenamefont {Watanabe}, \citenamefont {Taniguchi}, \citenamefont {Mak},\ and\ \citenamefont {Shan}}]{jin2020imaging}%
  \BibitemOpen
  \bibfield  {author} {\bibinfo {author} {\bibfnamefont {C.}~\bibnamefont {Jin}}, \bibinfo {author} {\bibfnamefont {Z.}~\bibnamefont {Tao}}, \bibinfo {author} {\bibfnamefont {K.}~\bibnamefont {Kang}}, \bibinfo {author} {\bibfnamefont {K.}~\bibnamefont {Watanabe}}, \bibinfo {author} {\bibfnamefont {T.}~\bibnamefont {Taniguchi}}, \bibinfo {author} {\bibfnamefont {K.~F.}\ \bibnamefont {Mak}},\ and\ \bibinfo {author} {\bibfnamefont {J.}~\bibnamefont {Shan}},\ }\href {https://doi.org/10.1038/s41563-020-0706-8} {\bibfield  {journal} {\bibinfo  {journal} {Nat. Mater.}\ }\textbf {\bibinfo {volume} {19}},\ \bibinfo {pages} {1290} (\bibinfo {year} {2020})}\BibitemShut {NoStop}%
\bibitem [{\citenamefont {Kittel}\ and\ \citenamefont {Van~Vleck}(1960)}]{Kittel1960}%
  \BibitemOpen
  \bibfield  {author} {\bibinfo {author} {\bibfnamefont {C.}~\bibnamefont {Kittel}}\ and\ \bibinfo {author} {\bibfnamefont {J.~H.}\ \bibnamefont {Van~Vleck}},\ }\href {https://doi.org/10.1103/PhysRev.118.1231} {\bibfield  {journal} {\bibinfo  {journal} {Phys. Rev.}\ }\textbf {\bibinfo {volume} {118}},\ \bibinfo {pages} {1231} (\bibinfo {year} {1960})}\BibitemShut {NoStop}%
\bibitem [{\citenamefont {Callen}\ and\ \citenamefont {Callen}(1965)}]{Callen1965}%
  \BibitemOpen
  \bibfield  {author} {\bibinfo {author} {\bibfnamefont {E.}~\bibnamefont {Callen}}\ and\ \bibinfo {author} {\bibfnamefont {H.~B.}\ \bibnamefont {Callen}},\ }\href {https://doi.org/10.1103/PhysRev.139.A455} {\bibfield  {journal} {\bibinfo  {journal} {Phys. Rev.}\ }\textbf {\bibinfo {volume} {139}},\ \bibinfo {pages} {A455} (\bibinfo {year} {1965})}\BibitemShut {NoStop}%
\bibitem [{\citenamefont {Callen}(1968)}]{Callen1968}%
  \BibitemOpen
  \bibfield  {author} {\bibinfo {author} {\bibfnamefont {E.}~\bibnamefont {Callen}},\ }\href {https://doi.org/10.1063/1.2163507} {\bibfield  {journal} {\bibinfo  {journal} {Journal of Applied Physics}\ }\textbf {\bibinfo {volume} {39}},\ \bibinfo {pages} {519} (\bibinfo {year} {1968})}\BibitemShut {NoStop}%
\bibitem [{\citenamefont {Zapf}\ \emph {et~al.}(2008)\citenamefont {Zapf}, \citenamefont {Correa}, \citenamefont {Sengupta}, \citenamefont {Batista}, \citenamefont {Tsukamoto}, \citenamefont {Kawashima}, \citenamefont {Egan}, \citenamefont {Pantea}, \citenamefont {Migliori}, \citenamefont {Betts}, \citenamefont {Jaime},\ and\ \citenamefont {Paduan-Filho}}]{zapf2008}%
  \BibitemOpen
  \bibfield  {author} {\bibinfo {author} {\bibfnamefont {V.~S.}\ \bibnamefont {Zapf}}, \bibinfo {author} {\bibfnamefont {V.~F.}\ \bibnamefont {Correa}}, \bibinfo {author} {\bibfnamefont {P.}~\bibnamefont {Sengupta}}, \bibinfo {author} {\bibfnamefont {C.~D.}\ \bibnamefont {Batista}}, \bibinfo {author} {\bibfnamefont {M.}~\bibnamefont {Tsukamoto}}, \bibinfo {author} {\bibfnamefont {N.}~\bibnamefont {Kawashima}}, \bibinfo {author} {\bibfnamefont {P.}~\bibnamefont {Egan}}, \bibinfo {author} {\bibfnamefont {C.}~\bibnamefont {Pantea}}, \bibinfo {author} {\bibfnamefont {A.}~\bibnamefont {Migliori}}, \bibinfo {author} {\bibfnamefont {J.~B.}\ \bibnamefont {Betts}}, \bibinfo {author} {\bibfnamefont {M.}~\bibnamefont {Jaime}},\ and\ \bibinfo {author} {\bibfnamefont {A.}~\bibnamefont {Paduan-Filho}},\ }\href {https://doi.org/10.1103/PhysRevB.77.020404} {\bibfield  {journal} {\bibinfo  {journal} {Phys. Rev. B}\ }\textbf {\bibinfo {volume} {77}},\ \bibinfo {pages} {020404} (\bibinfo {year} {2008})}\BibitemShut {NoStop}%
\bibitem [{\citenamefont {Küchler}\ \emph {et~al.}(2012)\citenamefont {Küchler}, \citenamefont {Bauer}, \citenamefont {Brando},\ and\ \citenamefont {Steglich}}]{kuchler2012}%
  \BibitemOpen
  \bibfield  {author} {\bibinfo {author} {\bibfnamefont {R.}~\bibnamefont {Küchler}}, \bibinfo {author} {\bibfnamefont {T.}~\bibnamefont {Bauer}}, \bibinfo {author} {\bibfnamefont {M.}~\bibnamefont {Brando}},\ and\ \bibinfo {author} {\bibfnamefont {F.}~\bibnamefont {Steglich}},\ }\href {https://doi.org/10.1063/1.4748864} {\bibfield  {journal} {\bibinfo  {journal} {Rev. Sci. Instrum.}\ }\textbf {\bibinfo {volume} {83}},\ \bibinfo {pages} {095102} (\bibinfo {year} {2012})}\BibitemShut {NoStop}%
\bibitem [{\citenamefont {Daou}\ \emph {et~al.}(2010)\citenamefont {Daou}, \citenamefont {Weickert}, \citenamefont {Nicklas}, \citenamefont {Steglich}, \citenamefont {Haase},\ and\ \citenamefont {Doerr}}]{Daou2010}%
  \BibitemOpen
  \bibfield  {author} {\bibinfo {author} {\bibfnamefont {R.}~\bibnamefont {Daou}}, \bibinfo {author} {\bibfnamefont {F.}~\bibnamefont {Weickert}}, \bibinfo {author} {\bibfnamefont {M.}~\bibnamefont {Nicklas}}, \bibinfo {author} {\bibfnamefont {F.}~\bibnamefont {Steglich}}, \bibinfo {author} {\bibfnamefont {A.}~\bibnamefont {Haase}},\ and\ \bibinfo {author} {\bibfnamefont {M.}~\bibnamefont {Doerr}},\ }\href {https://doi.org/10.1063/1.3356980} {\bibfield  {journal} {\bibinfo  {journal} {Rev. Sci. Instrum.}\ }\textbf {\bibinfo {volume} {81}},\ \bibinfo {pages} {033909} (\bibinfo {year} {2010})}\BibitemShut {NoStop}%
\bibitem [{\citenamefont {Park}\ \emph {et~al.}(2009)\citenamefont {Park}, \citenamefont {Graf}, \citenamefont {Murphy}, \citenamefont {Schmiedeshoff},\ and\ \citenamefont {Tozer}}]{Park2009}%
  \BibitemOpen
  \bibfield  {author} {\bibinfo {author} {\bibfnamefont {J.-H.}\ \bibnamefont {Park}}, \bibinfo {author} {\bibfnamefont {D.}~\bibnamefont {Graf}}, \bibinfo {author} {\bibfnamefont {T.~P.}\ \bibnamefont {Murphy}}, \bibinfo {author} {\bibfnamefont {G.~M.}\ \bibnamefont {Schmiedeshoff}},\ and\ \bibinfo {author} {\bibfnamefont {S.~W.}\ \bibnamefont {Tozer}},\ }\href {https://doi.org/10.1063/1.3258143} {\bibfield  {journal} {\bibinfo  {journal} {Rev. Sci. Instrum.}\ }\textbf {\bibinfo {volume} {80}},\ \bibinfo {pages} {116101} (\bibinfo {year} {2009})}\BibitemShut {NoStop}%
\bibitem [{\citenamefont {Ikeda}\ \emph {et~al.}(2019)\citenamefont {Ikeda}, \citenamefont {Furukawa}, \citenamefont {Janson}, \citenamefont {Matsuda}, \citenamefont {Takeyama}, \citenamefont {Yajima}, \citenamefont {Hiroi},\ and\ \citenamefont {Ishikawa}}]{Ikeda2019}%
  \BibitemOpen
  \bibfield  {author} {\bibinfo {author} {\bibfnamefont {A.}~\bibnamefont {Ikeda}}, \bibinfo {author} {\bibfnamefont {S.}~\bibnamefont {Furukawa}}, \bibinfo {author} {\bibfnamefont {O.}~\bibnamefont {Janson}}, \bibinfo {author} {\bibfnamefont {Y.~H.}\ \bibnamefont {Matsuda}}, \bibinfo {author} {\bibfnamefont {S.}~\bibnamefont {Takeyama}}, \bibinfo {author} {\bibfnamefont {T.}~\bibnamefont {Yajima}}, \bibinfo {author} {\bibfnamefont {Z.}~\bibnamefont {Hiroi}},\ and\ \bibinfo {author} {\bibfnamefont {H.}~\bibnamefont {Ishikawa}},\ }\href {https://doi.org/10.1103/PhysRevB.99.140412} {\bibfield  {journal} {\bibinfo  {journal} {Phys. Rev. B}\ }\textbf {\bibinfo {volume} {99}},\ \bibinfo {pages} {140412} (\bibinfo {year} {2019})}\BibitemShut {NoStop}%
\bibitem [{\citenamefont {Miyata}\ \emph {et~al.}(2021)\citenamefont {Miyata}, \citenamefont {Hikihara}, \citenamefont {Furukawa}, \citenamefont {Kremer}, \citenamefont {Zherlitsyn},\ and\ \citenamefont {Wosnitza}}]{Miyata2021}%
  \BibitemOpen
  \bibfield  {author} {\bibinfo {author} {\bibfnamefont {A.}~\bibnamefont {Miyata}}, \bibinfo {author} {\bibfnamefont {T.}~\bibnamefont {Hikihara}}, \bibinfo {author} {\bibfnamefont {S.}~\bibnamefont {Furukawa}}, \bibinfo {author} {\bibfnamefont {R.~K.}\ \bibnamefont {Kremer}}, \bibinfo {author} {\bibfnamefont {S.}~\bibnamefont {Zherlitsyn}},\ and\ \bibinfo {author} {\bibfnamefont {J.}~\bibnamefont {Wosnitza}},\ }\href {https://doi.org/10.1103/PhysRevB.103.014411} {\bibfield  {journal} {\bibinfo  {journal} {Phys. Rev. B}\ }\textbf {\bibinfo {volume} {103}},\ \bibinfo {pages} {014411} (\bibinfo {year} {2021})}\BibitemShut {NoStop}%
\bibitem [{\citenamefont {Chai}\ \emph {et~al.}(2021)\citenamefont {Chai}, \citenamefont {Lu}, \citenamefont {Du}, \citenamefont {Shen}, \citenamefont {Ma}, \citenamefont {Zhai}, \citenamefont {Wang}, \citenamefont {Shi}, \citenamefont {Li}, \citenamefont {Wang},\ and\ \citenamefont {Sun}}]{Chai2021}%
  \BibitemOpen
  \bibfield  {author} {\bibinfo {author} {\bibfnamefont {Y.}~\bibnamefont {Chai}}, \bibinfo {author} {\bibfnamefont {P.}~\bibnamefont {Lu}}, \bibinfo {author} {\bibfnamefont {H.}~\bibnamefont {Du}}, \bibinfo {author} {\bibfnamefont {J.}~\bibnamefont {Shen}}, \bibinfo {author} {\bibfnamefont {Y.}~\bibnamefont {Ma}}, \bibinfo {author} {\bibfnamefont {K.}~\bibnamefont {Zhai}}, \bibinfo {author} {\bibfnamefont {L.}~\bibnamefont {Wang}}, \bibinfo {author} {\bibfnamefont {Y.}~\bibnamefont {Shi}}, \bibinfo {author} {\bibfnamefont {H.}~\bibnamefont {Li}}, \bibinfo {author} {\bibfnamefont {W.}~\bibnamefont {Wang}},\ and\ \bibinfo {author} {\bibfnamefont {Y.}~\bibnamefont {Sun}},\ }\href {https://doi.org/10.1103/PhysRevB.104.L100413} {\bibfield  {journal} {\bibinfo  {journal} {Phys. Rev. B}\ }\textbf {\bibinfo {volume} {104}},\ \bibinfo {pages} {L100413} (\bibinfo {year} {2021})}\BibitemShut {NoStop}%
\bibitem [{\citenamefont {Zhang}\ \emph {et~al.}(2023{\natexlab{a}})\citenamefont {Zhang}, \citenamefont {Li}, \citenamefont {Zhang}, \citenamefont {Cao}, \citenamefont {Zhang}, \citenamefont {Li}, \citenamefont {Liu}, \citenamefont {Zhou}, \citenamefont {Sun}, \citenamefont {Wang},\ and\ \citenamefont {Chai}}]{zhang2023}%
  \BibitemOpen
  \bibfield  {author} {\bibinfo {author} {\bibfnamefont {Y.}~\bibnamefont {Zhang}}, \bibinfo {author} {\bibfnamefont {Z.}~\bibnamefont {Li}}, \bibinfo {author} {\bibfnamefont {J.}~\bibnamefont {Zhang}}, \bibinfo {author} {\bibfnamefont {N.}~\bibnamefont {Cao}}, \bibinfo {author} {\bibfnamefont {L.}~\bibnamefont {Zhang}}, \bibinfo {author} {\bibfnamefont {Y.}~\bibnamefont {Li}}, \bibinfo {author} {\bibfnamefont {S.}~\bibnamefont {Liu}}, \bibinfo {author} {\bibfnamefont {X.}~\bibnamefont {Zhou}}, \bibinfo {author} {\bibfnamefont {Y.}~\bibnamefont {Sun}}, \bibinfo {author} {\bibfnamefont {W.}~\bibnamefont {Wang}},\ and\ \bibinfo {author} {\bibfnamefont {Y.}~\bibnamefont {Chai}},\ }\href {https://doi.org/10.1103/PhysRevB.107.134417} {\bibfield  {journal} {\bibinfo  {journal} {Phys. Rev. B}\ }\textbf {\bibinfo {volume} {107}},\ \bibinfo {pages} {134417} (\bibinfo {year} {2023}{\natexlab{a}})}\BibitemShut {NoStop}%
\bibitem [{\citenamefont {Zhang}\ \emph {et~al.}(1995)\citenamefont {Zhang}, \citenamefont {Fung},\ and\ \citenamefont {Zeng}}]{Zhang_disspation}%
  \BibitemOpen
  \bibfield  {author} {\bibinfo {author} {\bibfnamefont {J.~X.}\ \bibnamefont {Zhang}}, \bibinfo {author} {\bibfnamefont {P.~C.~W.}\ \bibnamefont {Fung}},\ and\ \bibinfo {author} {\bibfnamefont {W.~G.}\ \bibnamefont {Zeng}},\ }\href {https://doi.org/10.1103/PhysRevB.52.268} {\bibfield  {journal} {\bibinfo  {journal} {Phys. Rev. B}\ }\textbf {\bibinfo {volume} {52}},\ \bibinfo {pages} {268} (\bibinfo {year} {1995})}\BibitemShut {NoStop}%
\bibitem [{\citenamefont {Topping}\ and\ \citenamefont {Blundell}(2018)}]{Topping_AC}%
  \BibitemOpen
  \bibfield  {author} {\bibinfo {author} {\bibfnamefont {C.~V.}\ \bibnamefont {Topping}}\ and\ \bibinfo {author} {\bibfnamefont {S.~J.}\ \bibnamefont {Blundell}},\ }\href {https://doi.org/10.1088/1361-648X/aaed96} {\bibfield  {journal} {\bibinfo  {journal} {J. Phys.: Condens. Matter}\ }\textbf {\bibinfo {volume} {31}},\ \bibinfo {pages} {013001} (\bibinfo {year} {2018})}\BibitemShut {NoStop}%
\bibitem [{\citenamefont {Kitaev}(2006)}]{KITAEV20062}%
  \BibitemOpen
  \bibfield  {author} {\bibinfo {author} {\bibfnamefont {A.}~\bibnamefont {Kitaev}},\ }\href {https://doi.org/https://doi.org/10.1016/j.aop.2005.10.005} {\bibfield  {journal} {\bibinfo  {journal} {Ann. Phys.}\ }\textbf {\bibinfo {volume} {321}},\ \bibinfo {pages} {2} (\bibinfo {year} {2006})},\ \bibinfo {note} {january Special Issue}\BibitemShut {NoStop}%
\bibitem [{\citenamefont {Kitaev}(2003)}]{KITAEV20032}%
  \BibitemOpen
  \bibfield  {author} {\bibinfo {author} {\bibfnamefont {A.}~\bibnamefont {Kitaev}},\ }\href {https://doi.org/https://doi.org/10.1016/S0003-4916(02)00018-0} {\bibfield  {journal} {\bibinfo  {journal} {Ann. Phys.}\ }\textbf {\bibinfo {volume} {303}},\ \bibinfo {pages} {2} (\bibinfo {year} {2003})}\BibitemShut {NoStop}%
\bibitem [{\citenamefont {Singh}\ \emph {et~al.}(2012)\citenamefont {Singh}, \citenamefont {Manni}, \citenamefont {Reuther}, \citenamefont {Berlijn}, \citenamefont {Thomale}, \citenamefont {Ku}, \citenamefont {Trebst},\ and\ \citenamefont {Gegenwart}}]{PhysRevLett.108.127203}%
  \BibitemOpen
  \bibfield  {author} {\bibinfo {author} {\bibfnamefont {Y.}~\bibnamefont {Singh}}, \bibinfo {author} {\bibfnamefont {S.}~\bibnamefont {Manni}}, \bibinfo {author} {\bibfnamefont {J.}~\bibnamefont {Reuther}}, \bibinfo {author} {\bibfnamefont {T.}~\bibnamefont {Berlijn}}, \bibinfo {author} {\bibfnamefont {R.}~\bibnamefont {Thomale}}, \bibinfo {author} {\bibfnamefont {W.}~\bibnamefont {Ku}}, \bibinfo {author} {\bibfnamefont {S.}~\bibnamefont {Trebst}},\ and\ \bibinfo {author} {\bibfnamefont {P.}~\bibnamefont {Gegenwart}},\ }\href {https://doi.org/10.1103/PhysRevLett.108.127203} {\bibfield  {journal} {\bibinfo  {journal} {Phys. Rev. Lett.}\ }\textbf {\bibinfo {volume} {108}},\ \bibinfo {pages} {127203} (\bibinfo {year} {2012})}\BibitemShut {NoStop}%
\bibitem [{\citenamefont {Chaloupka}\ \emph {et~al.}(2013)\citenamefont {Chaloupka}, \citenamefont {Jackeli},\ and\ \citenamefont {Khaliullin}}]{PhysRevLett.110.097204}%
  \BibitemOpen
  \bibfield  {author} {\bibinfo {author} {\bibfnamefont {J.~c.~v.}\ \bibnamefont {Chaloupka}}, \bibinfo {author} {\bibfnamefont {G.}~\bibnamefont {Jackeli}},\ and\ \bibinfo {author} {\bibfnamefont {G.}~\bibnamefont {Khaliullin}},\ }\href {https://doi.org/10.1103/PhysRevLett.110.097204} {\bibfield  {journal} {\bibinfo  {journal} {Phys. Rev. Lett.}\ }\textbf {\bibinfo {volume} {110}},\ \bibinfo {pages} {097204} (\bibinfo {year} {2013})}\BibitemShut {NoStop}%
\bibitem [{\citenamefont {Hwan~Chun}\ \emph {et~al.}(2015)\citenamefont {Hwan~Chun}, \citenamefont {Kim}, \citenamefont {Kim}, \citenamefont {Zheng}, \citenamefont {Stoumpos}, \citenamefont {Malliakas}, \citenamefont {Mitchell}, \citenamefont {Mehlawat}, \citenamefont {Singh}, \citenamefont {Choi} \emph {et~al.}}]{hwan2015direct}%
  \BibitemOpen
  \bibfield  {author} {\bibinfo {author} {\bibfnamefont {S.}~\bibnamefont {Hwan~Chun}}, \bibinfo {author} {\bibfnamefont {J.-W.}\ \bibnamefont {Kim}}, \bibinfo {author} {\bibfnamefont {J.}~\bibnamefont {Kim}}, \bibinfo {author} {\bibfnamefont {H.}~\bibnamefont {Zheng}}, \bibinfo {author} {\bibfnamefont {C.~C.}\ \bibnamefont {Stoumpos}}, \bibinfo {author} {\bibfnamefont {C.}~\bibnamefont {Malliakas}}, \bibinfo {author} {\bibfnamefont {J.}~\bibnamefont {Mitchell}}, \bibinfo {author} {\bibfnamefont {K.}~\bibnamefont {Mehlawat}}, \bibinfo {author} {\bibfnamefont {Y.}~\bibnamefont {Singh}}, \bibinfo {author} {\bibfnamefont {Y.}~\bibnamefont {Choi}}, \emph {et~al.},\ }\href {https://doi.org/10.1038/nphys3322} {\bibfield  {journal} {\bibinfo  {journal} {Nature Physics}\ }\textbf {\bibinfo {volume} {11}},\ \bibinfo {pages} {462} (\bibinfo {year} {2015})}\BibitemShut {NoStop}%
\bibitem [{\citenamefont {Singh}\ and\ \citenamefont {Gegenwart}(2010)}]{PhysRevB.82.064412}%
  \BibitemOpen
  \bibfield  {author} {\bibinfo {author} {\bibfnamefont {Y.}~\bibnamefont {Singh}}\ and\ \bibinfo {author} {\bibfnamefont {P.}~\bibnamefont {Gegenwart}},\ }\href {https://doi.org/10.1103/PhysRevB.82.064412} {\bibfield  {journal} {\bibinfo  {journal} {Phys. Rev. B}\ }\textbf {\bibinfo {volume} {82}},\ \bibinfo {pages} {064412} (\bibinfo {year} {2010})}\BibitemShut {NoStop}%
\bibitem [{\citenamefont {Banerjee}\ \emph {et~al.}(2016)\citenamefont {Banerjee}, \citenamefont {Bridges}, \citenamefont {Yan}, \citenamefont {Aczel}, \citenamefont {Li}, \citenamefont {Stone}, \citenamefont {Granroth}, \citenamefont {Lumsden}, \citenamefont {Yiu}, \citenamefont {Knolle} \emph {et~al.}}]{banerjee2016proximate}%
  \BibitemOpen
  \bibfield  {author} {\bibinfo {author} {\bibfnamefont {A.}~\bibnamefont {Banerjee}}, \bibinfo {author} {\bibfnamefont {C.}~\bibnamefont {Bridges}}, \bibinfo {author} {\bibfnamefont {J.-Q.}\ \bibnamefont {Yan}}, \bibinfo {author} {\bibfnamefont {A.}~\bibnamefont {Aczel}}, \bibinfo {author} {\bibfnamefont {L.}~\bibnamefont {Li}}, \bibinfo {author} {\bibfnamefont {M.}~\bibnamefont {Stone}}, \bibinfo {author} {\bibfnamefont {G.}~\bibnamefont {Granroth}}, \bibinfo {author} {\bibfnamefont {M.}~\bibnamefont {Lumsden}}, \bibinfo {author} {\bibfnamefont {Y.}~\bibnamefont {Yiu}}, \bibinfo {author} {\bibfnamefont {J.}~\bibnamefont {Knolle}}, \emph {et~al.},\ }\href {https://doi.org/10.1038/nmat4604} {\bibfield  {journal} {\bibinfo  {journal} {Nature materials}\ }\textbf {\bibinfo {volume} {15}},\ \bibinfo {pages} {733} (\bibinfo {year} {2016})}\BibitemShut {NoStop}%
\bibitem [{\citenamefont {Banerjee}\ \emph {et~al.}(2017)\citenamefont {Banerjee}, \citenamefont {Yan}, \citenamefont {Knolle}, \citenamefont {Bridges}, \citenamefont {Stone}, \citenamefont {Lumsden}, \citenamefont {Mandrus}, \citenamefont {Tennant}, \citenamefont {Moessner},\ and\ \citenamefont {Nagler}}]{Banerjee2017science}%
  \BibitemOpen
  \bibfield  {author} {\bibinfo {author} {\bibfnamefont {A.}~\bibnamefont {Banerjee}}, \bibinfo {author} {\bibfnamefont {J.}~\bibnamefont {Yan}}, \bibinfo {author} {\bibfnamefont {J.}~\bibnamefont {Knolle}}, \bibinfo {author} {\bibfnamefont {C.~A.}\ \bibnamefont {Bridges}}, \bibinfo {author} {\bibfnamefont {M.~B.}\ \bibnamefont {Stone}}, \bibinfo {author} {\bibfnamefont {M.~D.}\ \bibnamefont {Lumsden}}, \bibinfo {author} {\bibfnamefont {D.~G.}\ \bibnamefont {Mandrus}}, \bibinfo {author} {\bibfnamefont {D.~A.}\ \bibnamefont {Tennant}}, \bibinfo {author} {\bibfnamefont {R.}~\bibnamefont {Moessner}},\ and\ \bibinfo {author} {\bibfnamefont {S.~E.}\ \bibnamefont {Nagler}},\ }\href {https://doi.org/10.1126/science.aah6015} {\bibfield  {journal} {\bibinfo  {journal} {Science}\ }\textbf {\bibinfo {volume} {356}},\ \bibinfo {pages} {1055} (\bibinfo {year} {2017})},\ \Eprint {https://arxiv.org/abs/https://www.science.org/doi/pdf/10.1126/science.aah6015} {https://www.science.org/doi/pdf/10.1126/science.aah6015}
  \BibitemShut {NoStop}%
\bibitem [{\citenamefont {Plumb}\ \emph {et~al.}(2014)\citenamefont {Plumb}, \citenamefont {Clancy}, \citenamefont {Sandilands}, \citenamefont {Shankar}, \citenamefont {Hu}, \citenamefont {Burch}, \citenamefont {Kee},\ and\ \citenamefont {Kim}}]{PhysRevB.90.041112}%
  \BibitemOpen
  \bibfield  {author} {\bibinfo {author} {\bibfnamefont {K.~W.}\ \bibnamefont {Plumb}}, \bibinfo {author} {\bibfnamefont {J.~P.}\ \bibnamefont {Clancy}}, \bibinfo {author} {\bibfnamefont {L.~J.}\ \bibnamefont {Sandilands}}, \bibinfo {author} {\bibfnamefont {V.~V.}\ \bibnamefont {Shankar}}, \bibinfo {author} {\bibfnamefont {Y.~F.}\ \bibnamefont {Hu}}, \bibinfo {author} {\bibfnamefont {K.~S.}\ \bibnamefont {Burch}}, \bibinfo {author} {\bibfnamefont {H.-Y.}\ \bibnamefont {Kee}},\ and\ \bibinfo {author} {\bibfnamefont {Y.-J.}\ \bibnamefont {Kim}},\ }\href {https://doi.org/10.1103/PhysRevB.90.041112} {\bibfield  {journal} {\bibinfo  {journal} {Phys. Rev. B}\ }\textbf {\bibinfo {volume} {90}},\ \bibinfo {pages} {041112} (\bibinfo {year} {2014})}\BibitemShut {NoStop}%
\bibitem [{\citenamefont {Kasahara}\ \emph {et~al.}(2018)\citenamefont {Kasahara}, \citenamefont {Ohnishi}, \citenamefont {Mizukami}, \citenamefont {Tanaka}, \citenamefont {Ma}, \citenamefont {Sugii}, \citenamefont {Kurita}, \citenamefont {Tanaka}, \citenamefont {Nasu}, \citenamefont {Motome} \emph {et~al.}}]{kasahara2018majorana}%
  \BibitemOpen
  \bibfield  {author} {\bibinfo {author} {\bibfnamefont {Y.}~\bibnamefont {Kasahara}}, \bibinfo {author} {\bibfnamefont {T.}~\bibnamefont {Ohnishi}}, \bibinfo {author} {\bibfnamefont {Y.}~\bibnamefont {Mizukami}}, \bibinfo {author} {\bibfnamefont {O.}~\bibnamefont {Tanaka}}, \bibinfo {author} {\bibfnamefont {S.}~\bibnamefont {Ma}}, \bibinfo {author} {\bibfnamefont {K.}~\bibnamefont {Sugii}}, \bibinfo {author} {\bibfnamefont {N.}~\bibnamefont {Kurita}}, \bibinfo {author} {\bibfnamefont {H.}~\bibnamefont {Tanaka}}, \bibinfo {author} {\bibfnamefont {J.}~\bibnamefont {Nasu}}, \bibinfo {author} {\bibfnamefont {Y.}~\bibnamefont {Motome}}, \emph {et~al.},\ }\href {https://doi.org/10.1038/s41586-018-0274-0} {\bibfield  {journal} {\bibinfo  {journal} {Nature}\ }\textbf {\bibinfo {volume} {559}},\ \bibinfo {pages} {227} (\bibinfo {year} {2018})}\BibitemShut {NoStop}%
\bibitem [{\citenamefont {Shi}\ \emph {et~al.}(2021)\citenamefont {Shi}, \citenamefont {Wang}, \citenamefont {Zhong}, \citenamefont {Wang}, \citenamefont {Hu}, \citenamefont {Zhang}, \citenamefont {Liu}, \citenamefont {Dong}, \citenamefont {Wang},\ and\ \citenamefont {Wang}}]{PhysRevB.104.144408}%
  \BibitemOpen
  \bibfield  {author} {\bibinfo {author} {\bibfnamefont {L.~Y.}\ \bibnamefont {Shi}}, \bibinfo {author} {\bibfnamefont {X.~M.}\ \bibnamefont {Wang}}, \bibinfo {author} {\bibfnamefont {R.~D.}\ \bibnamefont {Zhong}}, \bibinfo {author} {\bibfnamefont {Z.~X.}\ \bibnamefont {Wang}}, \bibinfo {author} {\bibfnamefont {T.~C.}\ \bibnamefont {Hu}}, \bibinfo {author} {\bibfnamefont {S.~J.}\ \bibnamefont {Zhang}}, \bibinfo {author} {\bibfnamefont {Q.~M.}\ \bibnamefont {Liu}}, \bibinfo {author} {\bibfnamefont {T.}~\bibnamefont {Dong}}, \bibinfo {author} {\bibfnamefont {F.}~\bibnamefont {Wang}},\ and\ \bibinfo {author} {\bibfnamefont {N.~L.}\ \bibnamefont {Wang}},\ }\href {https://doi.org/10.1103/PhysRevB.104.144408} {\bibfield  {journal} {\bibinfo  {journal} {Phys. Rev. B}\ }\textbf {\bibinfo {volume} {104}},\ \bibinfo {pages} {144408} (\bibinfo {year} {2021})}\BibitemShut {NoStop}%
\bibitem [{\citenamefont {Zhong}\ \emph {et~al.}(2020)\citenamefont {Zhong}, \citenamefont {Gao}, \citenamefont {Ong},\ and\ \citenamefont {Cava}}]{ruidansciadv.aay6953}%
  \BibitemOpen
  \bibfield  {author} {\bibinfo {author} {\bibfnamefont {R.}~\bibnamefont {Zhong}}, \bibinfo {author} {\bibfnamefont {T.}~\bibnamefont {Gao}}, \bibinfo {author} {\bibfnamefont {N.~P.}\ \bibnamefont {Ong}},\ and\ \bibinfo {author} {\bibfnamefont {R.~J.}\ \bibnamefont {Cava}},\ }\href {https://doi.org/10.1126/sciadv.aay6953} {\bibfield  {journal} {\bibinfo  {journal} {Science Advances}\ }\textbf {\bibinfo {volume} {6}},\ \bibinfo {pages} {eaay6953} (\bibinfo {year} {2020})}\BibitemShut {NoStop}%
\bibitem [{\citenamefont {Zhang}\ \emph {et~al.}(2023{\natexlab{b}})\citenamefont {Zhang}, \citenamefont {Xu}, \citenamefont {Halloran}, \citenamefont {Zhong}, \citenamefont {Broholm}, \citenamefont {Cava}, \citenamefont {Drichko},\ and\ \citenamefont {Armitage}}]{zhang2023magnetic}%
  \BibitemOpen
  \bibfield  {author} {\bibinfo {author} {\bibfnamefont {X.}~\bibnamefont {Zhang}}, \bibinfo {author} {\bibfnamefont {Y.}~\bibnamefont {Xu}}, \bibinfo {author} {\bibfnamefont {T.}~\bibnamefont {Halloran}}, \bibinfo {author} {\bibfnamefont {R.}~\bibnamefont {Zhong}}, \bibinfo {author} {\bibfnamefont {C.}~\bibnamefont {Broholm}}, \bibinfo {author} {\bibfnamefont {R.}~\bibnamefont {Cava}}, \bibinfo {author} {\bibfnamefont {N.}~\bibnamefont {Drichko}},\ and\ \bibinfo {author} {\bibfnamefont {N.}~\bibnamefont {Armitage}},\ }\href {https://doi.org/10.1038/s41563-022-01403-1} {\bibfield  {journal} {\bibinfo  {journal} {Nature Materials}\ }\textbf {\bibinfo {volume} {22}},\ \bibinfo {pages} {58} (\bibinfo {year} {2023}{\natexlab{b}})}\BibitemShut {NoStop}%
\bibitem [{\citenamefont {Viciu}\ \emph {et~al.}(2007)\citenamefont {Viciu}, \citenamefont {Huang}, \citenamefont {Morosan}, \citenamefont {Zandbergen}, \citenamefont {Greenbaum}, \citenamefont {McQueen},\ and\ \citenamefont {Cava}}]{VICIU20071060}%
  \BibitemOpen
  \bibfield  {author} {\bibinfo {author} {\bibfnamefont {L.}~\bibnamefont {Viciu}}, \bibinfo {author} {\bibfnamefont {Q.}~\bibnamefont {Huang}}, \bibinfo {author} {\bibfnamefont {E.}~\bibnamefont {Morosan}}, \bibinfo {author} {\bibfnamefont {H.}~\bibnamefont {Zandbergen}}, \bibinfo {author} {\bibfnamefont {N.}~\bibnamefont {Greenbaum}}, \bibinfo {author} {\bibfnamefont {T.}~\bibnamefont {McQueen}},\ and\ \bibinfo {author} {\bibfnamefont {R.}~\bibnamefont {Cava}},\ }\href {https://doi.org/https://doi.org/10.1016/j.jssc.2007.01.002} {\bibfield  {journal} {\bibinfo  {journal} {J. Solid State Chem.}\ }\textbf {\bibinfo {volume} {180}},\ \bibinfo {pages} {1060} (\bibinfo {year} {2007})}\BibitemShut {NoStop}%
\bibitem [{\citenamefont {Chen}\ \emph {et~al.}(2021)\citenamefont {Chen}, \citenamefont {Li}, \citenamefont {Hu}, \citenamefont {Hu}, \citenamefont {Yue}, \citenamefont {Sutarto}, \citenamefont {He}, \citenamefont {Iida}, \citenamefont {Kamazawa}, \citenamefont {Yu}, \citenamefont {Lin},\ and\ \citenamefont {Li}}]{Li.L180404}%
  \BibitemOpen
  \bibfield  {author} {\bibinfo {author} {\bibfnamefont {W.}~\bibnamefont {Chen}}, \bibinfo {author} {\bibfnamefont {X.}~\bibnamefont {Li}}, \bibinfo {author} {\bibfnamefont {Z.}~\bibnamefont {Hu}}, \bibinfo {author} {\bibfnamefont {Z.}~\bibnamefont {Hu}}, \bibinfo {author} {\bibfnamefont {L.}~\bibnamefont {Yue}}, \bibinfo {author} {\bibfnamefont {R.}~\bibnamefont {Sutarto}}, \bibinfo {author} {\bibfnamefont {F.}~\bibnamefont {He}}, \bibinfo {author} {\bibfnamefont {K.}~\bibnamefont {Iida}}, \bibinfo {author} {\bibfnamefont {K.}~\bibnamefont {Kamazawa}}, \bibinfo {author} {\bibfnamefont {W.}~\bibnamefont {Yu}}, \bibinfo {author} {\bibfnamefont {X.}~\bibnamefont {Lin}},\ and\ \bibinfo {author} {\bibfnamefont {Y.}~\bibnamefont {Li}},\ }\href {https://doi.org/10.1103/PhysRevB.103.L180404} {\bibfield  {journal} {\bibinfo  {journal} {Phys. Rev. B}\ }\textbf {\bibinfo {volume} {103}},\ \bibinfo {pages} {L180404} (\bibinfo {year} {2021})}\BibitemShut {NoStop}%
\bibitem [{\citenamefont {Lin}\ \emph {et~al.}(2021)\citenamefont {Lin}, \citenamefont {Jeong}, \citenamefont {Kim}, \citenamefont {Wang}, \citenamefont {Huang}, \citenamefont {Masuda}, \citenamefont {Asai}, \citenamefont {Itoh}, \citenamefont {G{\"u}nther}, \citenamefont {Russina} \emph {et~al.}}]{lin2021field}%
  \BibitemOpen
  \bibfield  {author} {\bibinfo {author} {\bibfnamefont {G.}~\bibnamefont {Lin}}, \bibinfo {author} {\bibfnamefont {J.}~\bibnamefont {Jeong}}, \bibinfo {author} {\bibfnamefont {C.}~\bibnamefont {Kim}}, \bibinfo {author} {\bibfnamefont {Y.}~\bibnamefont {Wang}}, \bibinfo {author} {\bibfnamefont {Q.}~\bibnamefont {Huang}}, \bibinfo {author} {\bibfnamefont {T.}~\bibnamefont {Masuda}}, \bibinfo {author} {\bibfnamefont {S.}~\bibnamefont {Asai}}, \bibinfo {author} {\bibfnamefont {S.}~\bibnamefont {Itoh}}, \bibinfo {author} {\bibfnamefont {G.}~\bibnamefont {G{\"u}nther}}, \bibinfo {author} {\bibfnamefont {M.}~\bibnamefont {Russina}}, \emph {et~al.},\ }\href {https://doi.org/10.1038/s41467-021-25567-7} {\bibfield  {journal} {\bibinfo  {journal} {Nature communications}\ }\textbf {\bibinfo {volume} {12}},\ \bibinfo {pages} {5559} (\bibinfo {year} {2021})}\BibitemShut {NoStop}%
\bibitem [{\citenamefont {Songvilay}\ \emph {et~al.}(2020{\natexlab{a}})\citenamefont {Songvilay}, \citenamefont {Robert}, \citenamefont {Petit}, \citenamefont {Rodriguez-Rivera}, \citenamefont {Ratcliff}, \citenamefont {Damay}, \citenamefont {Bal\'edent}, \citenamefont {Jim\'enez-Ruiz}, \citenamefont {Lejay}, \citenamefont {Pachoud}, \citenamefont {Hadj-Azzem}, \citenamefont {Simonet},\ and\ \citenamefont {Stock}}]{Songvilay224429}%
  \BibitemOpen
  \bibfield  {author} {\bibinfo {author} {\bibfnamefont {M.}~\bibnamefont {Songvilay}}, \bibinfo {author} {\bibfnamefont {J.}~\bibnamefont {Robert}}, \bibinfo {author} {\bibfnamefont {S.}~\bibnamefont {Petit}}, \bibinfo {author} {\bibfnamefont {J.~A.}\ \bibnamefont {Rodriguez-Rivera}}, \bibinfo {author} {\bibfnamefont {W.~D.}\ \bibnamefont {Ratcliff}}, \bibinfo {author} {\bibfnamefont {F.}~\bibnamefont {Damay}}, \bibinfo {author} {\bibfnamefont {V.}~\bibnamefont {Bal\'edent}}, \bibinfo {author} {\bibfnamefont {M.}~\bibnamefont {Jim\'enez-Ruiz}}, \bibinfo {author} {\bibfnamefont {P.}~\bibnamefont {Lejay}}, \bibinfo {author} {\bibfnamefont {E.}~\bibnamefont {Pachoud}}, \bibinfo {author} {\bibfnamefont {A.}~\bibnamefont {Hadj-Azzem}}, \bibinfo {author} {\bibfnamefont {V.}~\bibnamefont {Simonet}},\ and\ \bibinfo {author} {\bibfnamefont {C.}~\bibnamefont {Stock}},\ }\href {https://doi.org/10.1103/PhysRevB.102.224429} {\bibfield  {journal} {\bibinfo  {journal} {Phys. Rev. B}\ }\textbf {\bibinfo {volume} {102}},\
  \bibinfo {pages} {224429} (\bibinfo {year} {2020}{\natexlab{a}})}\BibitemShut {NoStop}%
\bibitem [{\citenamefont {Yan}\ \emph {et~al.}(2019{\natexlab{a}})\citenamefont {Yan}, \citenamefont {Okamoto}, \citenamefont {Wu}, \citenamefont {Zheng}, \citenamefont {Zhou}, \citenamefont {Cao},\ and\ \citenamefont {McGuire}}]{Yan074405}%
  \BibitemOpen
  \bibfield  {author} {\bibinfo {author} {\bibfnamefont {J.-Q.}\ \bibnamefont {Yan}}, \bibinfo {author} {\bibfnamefont {S.}~\bibnamefont {Okamoto}}, \bibinfo {author} {\bibfnamefont {Y.}~\bibnamefont {Wu}}, \bibinfo {author} {\bibfnamefont {Q.}~\bibnamefont {Zheng}}, \bibinfo {author} {\bibfnamefont {H.~D.}\ \bibnamefont {Zhou}}, \bibinfo {author} {\bibfnamefont {H.~B.}\ \bibnamefont {Cao}},\ and\ \bibinfo {author} {\bibfnamefont {M.~A.}\ \bibnamefont {McGuire}},\ }\href {https://doi.org/10.1103/PhysRevMaterials.3.074405} {\bibfield  {journal} {\bibinfo  {journal} {Phys. Rev. Mater.}\ }\textbf {\bibinfo {volume} {3}},\ \bibinfo {pages} {074405} (\bibinfo {year} {2019}{\natexlab{a}})}\BibitemShut {NoStop}%
\bibitem [{\citenamefont {Li}\ \emph {et~al.}(2022)\citenamefont {Li}, \citenamefont {Gu}, \citenamefont {Chen}, \citenamefont {Garlea}, \citenamefont {Iida}, \citenamefont {Kamazawa}, \citenamefont {Li}, \citenamefont {Deng}, \citenamefont {Xiao}, \citenamefont {Zheng}, \citenamefont {Ye}, \citenamefont {Peng}, \citenamefont {Zaliznyak}, \citenamefont {Tranquada},\ and\ \citenamefont {Li}}]{Li2022}%
  \BibitemOpen
  \bibfield  {author} {\bibinfo {author} {\bibfnamefont {X.}~\bibnamefont {Li}}, \bibinfo {author} {\bibfnamefont {Y.}~\bibnamefont {Gu}}, \bibinfo {author} {\bibfnamefont {Y.}~\bibnamefont {Chen}}, \bibinfo {author} {\bibfnamefont {V.~O.}\ \bibnamefont {Garlea}}, \bibinfo {author} {\bibfnamefont {K.}~\bibnamefont {Iida}}, \bibinfo {author} {\bibfnamefont {K.}~\bibnamefont {Kamazawa}}, \bibinfo {author} {\bibfnamefont {Y.}~\bibnamefont {Li}}, \bibinfo {author} {\bibfnamefont {G.}~\bibnamefont {Deng}}, \bibinfo {author} {\bibfnamefont {Q.}~\bibnamefont {Xiao}}, \bibinfo {author} {\bibfnamefont {X.}~\bibnamefont {Zheng}}, \bibinfo {author} {\bibfnamefont {Z.}~\bibnamefont {Ye}}, \bibinfo {author} {\bibfnamefont {Y.}~\bibnamefont {Peng}}, \bibinfo {author} {\bibfnamefont {I.~A.}\ \bibnamefont {Zaliznyak}}, \bibinfo {author} {\bibfnamefont {J.~M.}\ \bibnamefont {Tranquada}},\ and\ \bibinfo {author} {\bibfnamefont {Y.}~\bibnamefont {Li}},\ }\href {https://doi.org/10.1103/PhysRevX.12.041024} {\bibfield  {journal}
  {\bibinfo  {journal} {Phys. Rev. X}\ }\textbf {\bibinfo {volume} {12}},\ \bibinfo {pages} {041024} (\bibinfo {year} {2022})}\BibitemShut {NoStop}%
\bibitem [{\citenamefont {Vavilova}\ \emph {et~al.}(2023)\citenamefont {Vavilova}, \citenamefont {Vasilchikova}, \citenamefont {Vasiliev}, \citenamefont {Mikhailova}, \citenamefont {Nalbandyan}, \citenamefont {Zvereva},\ and\ \citenamefont {Streltsov}}]{Vavi2023}%
  \BibitemOpen
  \bibfield  {author} {\bibinfo {author} {\bibfnamefont {E.}~\bibnamefont {Vavilova}}, \bibinfo {author} {\bibfnamefont {T.}~\bibnamefont {Vasilchikova}}, \bibinfo {author} {\bibfnamefont {A.}~\bibnamefont {Vasiliev}}, \bibinfo {author} {\bibfnamefont {D.}~\bibnamefont {Mikhailova}}, \bibinfo {author} {\bibfnamefont {V.}~\bibnamefont {Nalbandyan}}, \bibinfo {author} {\bibfnamefont {E.}~\bibnamefont {Zvereva}},\ and\ \bibinfo {author} {\bibfnamefont {S.~V.}\ \bibnamefont {Streltsov}},\ }\href {https://doi.org/10.1103/PhysRevB.107.054411} {\bibfield  {journal} {\bibinfo  {journal} {Phys. Rev. B}\ }\textbf {\bibinfo {volume} {107}},\ \bibinfo {pages} {054411} (\bibinfo {year} {2023})}\BibitemShut {NoStop}%
\bibitem [{\citenamefont {Hu}\ \emph {et~al.}(2024)\citenamefont {Hu}, \citenamefont {Chen}, \citenamefont {Cui}, \citenamefont {Li}, \citenamefont {Li}, \citenamefont {Xu}, \citenamefont {Chen}, \citenamefont {Li}, \citenamefont {Gu}, \citenamefont {Yu}, \citenamefont {Zhou}, \citenamefont {Li},\ and\ \citenamefont {Yu}}]{Hu2024}%
  \BibitemOpen
  \bibfield  {author} {\bibinfo {author} {\bibfnamefont {Z.}~\bibnamefont {Hu}}, \bibinfo {author} {\bibfnamefont {Y.}~\bibnamefont {Chen}}, \bibinfo {author} {\bibfnamefont {Y.}~\bibnamefont {Cui}}, \bibinfo {author} {\bibfnamefont {S.}~\bibnamefont {Li}}, \bibinfo {author} {\bibfnamefont {C.}~\bibnamefont {Li}}, \bibinfo {author} {\bibfnamefont {X.}~\bibnamefont {Xu}}, \bibinfo {author} {\bibfnamefont {Y.}~\bibnamefont {Chen}}, \bibinfo {author} {\bibfnamefont {X.}~\bibnamefont {Li}}, \bibinfo {author} {\bibfnamefont {Y.}~\bibnamefont {Gu}}, \bibinfo {author} {\bibfnamefont {R.}~\bibnamefont {Yu}}, \bibinfo {author} {\bibfnamefont {R.}~\bibnamefont {Zhou}}, \bibinfo {author} {\bibfnamefont {Y.}~\bibnamefont {Li}},\ and\ \bibinfo {author} {\bibfnamefont {W.}~\bibnamefont {Yu}},\ }\href {https://doi.org/10.1103/PhysRevB.109.054411} {\bibfield  {journal} {\bibinfo  {journal} {Phys. Rev. B}\ }\textbf {\bibinfo {volume} {109}},\ \bibinfo {pages} {054411} (\bibinfo {year} {2024})}\BibitemShut {NoStop}%
\bibitem [{\citenamefont {Liu}\ \emph {et~al.}(2020)\citenamefont {Liu}, \citenamefont {Chaloupka},\ and\ \citenamefont {Khaliullin}}]{Liu2020}%
  \BibitemOpen
  \bibfield  {author} {\bibinfo {author} {\bibfnamefont {H.}~\bibnamefont {Liu}}, \bibinfo {author} {\bibfnamefont {J.~c.~v.}\ \bibnamefont {Chaloupka}},\ and\ \bibinfo {author} {\bibfnamefont {G.}~\bibnamefont {Khaliullin}},\ }\href {https://doi.org/10.1103/PhysRevLett.125.047201} {\bibfield  {journal} {\bibinfo  {journal} {Phys. Rev. Lett.}\ }\textbf {\bibinfo {volume} {125}},\ \bibinfo {pages} {047201} (\bibinfo {year} {2020})}\BibitemShut {NoStop}%
\bibitem [{\citenamefont {Songvilay}\ \emph {et~al.}(2020{\natexlab{b}})\citenamefont {Songvilay}, \citenamefont {Robert}, \citenamefont {Petit}, \citenamefont {Rodriguez-Rivera}, \citenamefont {Ratcliff}, \citenamefont {Damay}, \citenamefont {Bal\'edent}, \citenamefont {Jim\'enez-Ruiz}, \citenamefont {Lejay}, \citenamefont {Pachoud}, \citenamefont {Hadj-Azzem}, \citenamefont {Simonet},\ and\ \citenamefont {Stock}}]{Songvi2020}%
  \BibitemOpen
  \bibfield  {author} {\bibinfo {author} {\bibfnamefont {M.}~\bibnamefont {Songvilay}}, \bibinfo {author} {\bibfnamefont {J.}~\bibnamefont {Robert}}, \bibinfo {author} {\bibfnamefont {S.}~\bibnamefont {Petit}}, \bibinfo {author} {\bibfnamefont {J.~A.}\ \bibnamefont {Rodriguez-Rivera}}, \bibinfo {author} {\bibfnamefont {W.~D.}\ \bibnamefont {Ratcliff}}, \bibinfo {author} {\bibfnamefont {F.}~\bibnamefont {Damay}}, \bibinfo {author} {\bibfnamefont {V.}~\bibnamefont {Bal\'edent}}, \bibinfo {author} {\bibfnamefont {M.}~\bibnamefont {Jim\'enez-Ruiz}}, \bibinfo {author} {\bibfnamefont {P.}~\bibnamefont {Lejay}}, \bibinfo {author} {\bibfnamefont {E.}~\bibnamefont {Pachoud}}, \bibinfo {author} {\bibfnamefont {A.}~\bibnamefont {Hadj-Azzem}}, \bibinfo {author} {\bibfnamefont {V.}~\bibnamefont {Simonet}},\ and\ \bibinfo {author} {\bibfnamefont {C.}~\bibnamefont {Stock}},\ }\href {https://doi.org/10.1103/PhysRevB.102.224429} {\bibfield  {journal} {\bibinfo  {journal} {Phys. Rev. B}\ }\textbf {\bibinfo {volume} {102}},\
  \bibinfo {pages} {224429} (\bibinfo {year} {2020}{\natexlab{b}})}\BibitemShut {NoStop}%
\bibitem [{\citenamefont {Sanders}\ \emph {et~al.}(2022)\citenamefont {Sanders}, \citenamefont {Mole}, \citenamefont {Liu}, \citenamefont {Brown}, \citenamefont {Yu}, \citenamefont {Ling},\ and\ \citenamefont {Rachel}}]{Sanders2022}%
  \BibitemOpen
  \bibfield  {author} {\bibinfo {author} {\bibfnamefont {A.~L.}\ \bibnamefont {Sanders}}, \bibinfo {author} {\bibfnamefont {R.~A.}\ \bibnamefont {Mole}}, \bibinfo {author} {\bibfnamefont {J.}~\bibnamefont {Liu}}, \bibinfo {author} {\bibfnamefont {A.~J.}\ \bibnamefont {Brown}}, \bibinfo {author} {\bibfnamefont {D.}~\bibnamefont {Yu}}, \bibinfo {author} {\bibfnamefont {C.~D.}\ \bibnamefont {Ling}},\ and\ \bibinfo {author} {\bibfnamefont {S.}~\bibnamefont {Rachel}},\ }\href {https://doi.org/10.1103/PhysRevB.106.014413} {\bibfield  {journal} {\bibinfo  {journal} {Phys. Rev. B}\ }\textbf {\bibinfo {volume} {106}},\ \bibinfo {pages} {014413} (\bibinfo {year} {2022})}\BibitemShut {NoStop}%
\bibitem [{\citenamefont {Kim}\ \emph {et~al.}(2021)\citenamefont {Kim}, \citenamefont {Jeong}, \citenamefont {Lin}, \citenamefont {Park}, \citenamefont {Masuda}, \citenamefont {Asai}, \citenamefont {Itoh}, \citenamefont {Kim}, \citenamefont {Zhou}, \citenamefont {Ma},\ and\ \citenamefont {Park}}]{Kim_2022}%
  \BibitemOpen
  \bibfield  {author} {\bibinfo {author} {\bibfnamefont {C.}~\bibnamefont {Kim}}, \bibinfo {author} {\bibfnamefont {J.}~\bibnamefont {Jeong}}, \bibinfo {author} {\bibfnamefont {G.}~\bibnamefont {Lin}}, \bibinfo {author} {\bibfnamefont {P.}~\bibnamefont {Park}}, \bibinfo {author} {\bibfnamefont {T.}~\bibnamefont {Masuda}}, \bibinfo {author} {\bibfnamefont {S.}~\bibnamefont {Asai}}, \bibinfo {author} {\bibfnamefont {S.}~\bibnamefont {Itoh}}, \bibinfo {author} {\bibfnamefont {H.-S.}\ \bibnamefont {Kim}}, \bibinfo {author} {\bibfnamefont {H.}~\bibnamefont {Zhou}}, \bibinfo {author} {\bibfnamefont {J.}~\bibnamefont {Ma}},\ and\ \bibinfo {author} {\bibfnamefont {J.-G.}\ \bibnamefont {Park}},\ }\href {https://doi.org/10.1088/1361-648X/ac2644} {\bibfield  {journal} {\bibinfo  {journal} {J. Phys.: Condens. Matter}\ }\textbf {\bibinfo {volume} {34}},\ \bibinfo {pages} {045802} (\bibinfo {year} {2021})}\BibitemShut {NoStop}%
\bibitem [{\citenamefont {Wong}\ \emph {et~al.}(2016)\citenamefont {Wong}, \citenamefont {Avdeev},\ and\ \citenamefont {Ling}}]{WONG201618}%
  \BibitemOpen
  \bibfield  {author} {\bibinfo {author} {\bibfnamefont {C.}~\bibnamefont {Wong}}, \bibinfo {author} {\bibfnamefont {M.}~\bibnamefont {Avdeev}},\ and\ \bibinfo {author} {\bibfnamefont {C.~D.}\ \bibnamefont {Ling}},\ }\href {https://doi.org/https://doi.org/10.1016/j.jssc.2016.07.032} {\bibfield  {journal} {\bibinfo  {journal} {J. Solid State Chem.}\ }\textbf {\bibinfo {volume} {243}},\ \bibinfo {pages} {18} (\bibinfo {year} {2016})}\BibitemShut {NoStop}%
\bibitem [{\citenamefont {Yan}\ \emph {et~al.}(2019{\natexlab{b}})\citenamefont {Yan}, \citenamefont {Okamoto}, \citenamefont {Wu}, \citenamefont {Zheng}, \citenamefont {Zhou}, \citenamefont {Cao},\ and\ \citenamefont {McGuire}}]{Yan2019}%
  \BibitemOpen
  \bibfield  {author} {\bibinfo {author} {\bibfnamefont {J.-Q.}\ \bibnamefont {Yan}}, \bibinfo {author} {\bibfnamefont {S.}~\bibnamefont {Okamoto}}, \bibinfo {author} {\bibfnamefont {Y.}~\bibnamefont {Wu}}, \bibinfo {author} {\bibfnamefont {Q.}~\bibnamefont {Zheng}}, \bibinfo {author} {\bibfnamefont {H.~D.}\ \bibnamefont {Zhou}}, \bibinfo {author} {\bibfnamefont {H.~B.}\ \bibnamefont {Cao}},\ and\ \bibinfo {author} {\bibfnamefont {M.~A.}\ \bibnamefont {McGuire}},\ }\href {https://doi.org/10.1103/PhysRevMaterials.3.074405} {\bibfield  {journal} {\bibinfo  {journal} {Phys. Rev. Mater.}\ }\textbf {\bibinfo {volume} {3}},\ \bibinfo {pages} {074405} (\bibinfo {year} {2019}{\natexlab{b}})}\BibitemShut {NoStop}%
\bibitem [{\citenamefont {Stratan}\ \emph {et~al.}(2019)\citenamefont {Stratan}, \citenamefont {Shukaev}, \citenamefont {Vasilchikova}, \citenamefont {Vasiliev}, \citenamefont {Korshunov}, \citenamefont {Kurbakov}, \citenamefont {Nalbandyan},\ and\ \citenamefont {Zvereva}}]{C9NJ03627J}%
  \BibitemOpen
  \bibfield  {author} {\bibinfo {author} {\bibfnamefont {M.~I.}\ \bibnamefont {Stratan}}, \bibinfo {author} {\bibfnamefont {I.~L.}\ \bibnamefont {Shukaev}}, \bibinfo {author} {\bibfnamefont {T.~M.}\ \bibnamefont {Vasilchikova}}, \bibinfo {author} {\bibfnamefont {A.~N.}\ \bibnamefont {Vasiliev}}, \bibinfo {author} {\bibfnamefont {A.~N.}\ \bibnamefont {Korshunov}}, \bibinfo {author} {\bibfnamefont {A.~I.}\ \bibnamefont {Kurbakov}}, \bibinfo {author} {\bibfnamefont {V.~B.}\ \bibnamefont {Nalbandyan}},\ and\ \bibinfo {author} {\bibfnamefont {E.~A.}\ \bibnamefont {Zvereva}},\ }\href {https://doi.org/10.1039/C9NJ03627J} {\bibfield  {journal} {\bibinfo  {journal} {New J. Chem.}\ }\textbf {\bibinfo {volume} {43}},\ \bibinfo {pages} {13545} (\bibinfo {year} {2019})}\BibitemShut {NoStop}%
\bibitem [{\citenamefont {van Veenendaal}\ \emph {et~al.}(2023)\citenamefont {van Veenendaal}, \citenamefont {Poldi}, \citenamefont {Veiga}, \citenamefont {Bencok}, \citenamefont {Fabbris}, \citenamefont {Tartaglia}, \citenamefont {McChesney}, \citenamefont {Freeland}, \citenamefont {Hemley}, \citenamefont {Zheng}, \citenamefont {Mitchell}, \citenamefont {Yan},\ and\ \citenamefont {Haskel}}]{veenendaal2023}%
  \BibitemOpen
  \bibfield  {author} {\bibinfo {author} {\bibfnamefont {M.}~\bibnamefont {van Veenendaal}}, \bibinfo {author} {\bibfnamefont {E.~H.~T.}\ \bibnamefont {Poldi}}, \bibinfo {author} {\bibfnamefont {L.~S.~I.}\ \bibnamefont {Veiga}}, \bibinfo {author} {\bibfnamefont {P.}~\bibnamefont {Bencok}}, \bibinfo {author} {\bibfnamefont {G.}~\bibnamefont {Fabbris}}, \bibinfo {author} {\bibfnamefont {R.}~\bibnamefont {Tartaglia}}, \bibinfo {author} {\bibfnamefont {J.~L.}\ \bibnamefont {McChesney}}, \bibinfo {author} {\bibfnamefont {J.~W.}\ \bibnamefont {Freeland}}, \bibinfo {author} {\bibfnamefont {R.~J.}\ \bibnamefont {Hemley}}, \bibinfo {author} {\bibfnamefont {H.}~\bibnamefont {Zheng}}, \bibinfo {author} {\bibfnamefont {J.~F.}\ \bibnamefont {Mitchell}}, \bibinfo {author} {\bibfnamefont {J.-Q.}\ \bibnamefont {Yan}},\ and\ \bibinfo {author} {\bibfnamefont {D.}~\bibnamefont {Haskel}},\ }\href {https://doi.org/10.1103/PhysRevB.107.214443} {\bibfield  {journal} {\bibinfo  {journal} {Phys. Rev. B}\ }\textbf {\bibinfo {volume}
  {107}},\ \bibinfo {pages} {214443} (\bibinfo {year} {2023})}\BibitemShut {NoStop}%
\bibitem [{sup()}]{supplemental}%
  \BibitemOpen
  \href@noop {} {\bibinfo {title} {See supplemental material for crystal characterizations and more experimental data.}}\BibitemShut {Stop}%
\bibitem [{\citenamefont {Guillou}\ \emph {et~al.}(2018)\citenamefont {Guillou}, \citenamefont {Pathak}, \citenamefont {Paudyal}, \citenamefont {Mudryk}, \citenamefont {Wilhelm}, \citenamefont {Rogalev},\ and\ \citenamefont {Pecharsky}}]{Guillou_first}%
  \BibitemOpen
  \bibfield  {author} {\bibinfo {author} {\bibfnamefont {F.}~\bibnamefont {Guillou}}, \bibinfo {author} {\bibfnamefont {A.~K.}\ \bibnamefont {Pathak}}, \bibinfo {author} {\bibfnamefont {D.}~\bibnamefont {Paudyal}}, \bibinfo {author} {\bibfnamefont {Y.}~\bibnamefont {Mudryk}}, \bibinfo {author} {\bibfnamefont {F.}~\bibnamefont {Wilhelm}}, \bibinfo {author} {\bibfnamefont {A.}~\bibnamefont {Rogalev}},\ and\ \bibinfo {author} {\bibfnamefont {V.~K.}\ \bibnamefont {Pecharsky}},\ }\href {https://doi.org/10.1038/s41467-018-05268-4} {\bibfield  {journal} {\bibinfo  {journal} {Nature Communications}\ }\textbf {\bibinfo {volume} {9}},\ \bibinfo {pages} {2925} (\bibinfo {year} {2018})}\BibitemShut {NoStop}%
\bibitem [{\citenamefont {Alho}\ \emph {et~al.}(2020)\citenamefont {Alho}, \citenamefont {Ribeiro}, \citenamefont {von Ranke}, \citenamefont {Guillou}, \citenamefont {Mudryk},\ and\ \citenamefont {Pecharsky}}]{Alho_first}%
  \BibitemOpen
  \bibfield  {author} {\bibinfo {author} {\bibfnamefont {B.~P.}\ \bibnamefont {Alho}}, \bibinfo {author} {\bibfnamefont {P.~O.}\ \bibnamefont {Ribeiro}}, \bibinfo {author} {\bibfnamefont {P.~J.}\ \bibnamefont {von Ranke}}, \bibinfo {author} {\bibfnamefont {F.}~\bibnamefont {Guillou}}, \bibinfo {author} {\bibfnamefont {Y.}~\bibnamefont {Mudryk}},\ and\ \bibinfo {author} {\bibfnamefont {V.~K.}\ \bibnamefont {Pecharsky}},\ }\href {https://doi.org/10.1103/PhysRevB.102.134425} {\bibfield  {journal} {\bibinfo  {journal} {Phys. Rev. B}\ }\textbf {\bibinfo {volume} {102}},\ \bibinfo {pages} {134425} (\bibinfo {year} {2020})}\BibitemShut {NoStop}%
\bibitem [{\citenamefont {Biswas}\ \emph {et~al.}(2020)\citenamefont {Biswas}, \citenamefont {Zarkevich}, \citenamefont {Pathak}, \citenamefont {Dolotko}, \citenamefont {Hlova}, \citenamefont {Smirnov}, \citenamefont {Mudryk}, \citenamefont {Johnson},\ and\ \citenamefont {Pecharsky}}]{Biswas_first}%
  \BibitemOpen
  \bibfield  {author} {\bibinfo {author} {\bibfnamefont {A.}~\bibnamefont {Biswas}}, \bibinfo {author} {\bibfnamefont {N.~A.}\ \bibnamefont {Zarkevich}}, \bibinfo {author} {\bibfnamefont {A.~K.}\ \bibnamefont {Pathak}}, \bibinfo {author} {\bibfnamefont {O.}~\bibnamefont {Dolotko}}, \bibinfo {author} {\bibfnamefont {I.~Z.}\ \bibnamefont {Hlova}}, \bibinfo {author} {\bibfnamefont {A.~V.}\ \bibnamefont {Smirnov}}, \bibinfo {author} {\bibfnamefont {Y.}~\bibnamefont {Mudryk}}, \bibinfo {author} {\bibfnamefont {D.~D.}\ \bibnamefont {Johnson}},\ and\ \bibinfo {author} {\bibfnamefont {V.~K.}\ \bibnamefont {Pecharsky}},\ }\href {https://doi.org/10.1103/PhysRevB.101.224402} {\bibfield  {journal} {\bibinfo  {journal} {Phys. Rev. B}\ }\textbf {\bibinfo {volume} {101}},\ \bibinfo {pages} {224402} (\bibinfo {year} {2020})}\BibitemShut {NoStop}%
\bibitem [{Fan()}]{Fan_kappa}%
  \BibitemOpen
  \href@noop {} {\bibinfo {title} {H. fan et al., unpublished.}}\BibitemShut {Stop}%
\bibitem [{\citenamefont {Grigera}\ \emph {et~al.}(2001)\citenamefont {Grigera}, \citenamefont {Perry}, \citenamefont {Schofield}, \citenamefont {Chiao}, \citenamefont {Julian}, \citenamefont {Lonzarich}, \citenamefont {Ikeda}, \citenamefont {Maeno}, \citenamefont {Millis},\ and\ \citenamefont {Mackenzie}}]{SrRuO}%
  \BibitemOpen
  \bibfield  {author} {\bibinfo {author} {\bibfnamefont {S.~A.}\ \bibnamefont {Grigera}}, \bibinfo {author} {\bibfnamefont {R.~S.}\ \bibnamefont {Perry}}, \bibinfo {author} {\bibfnamefont {A.~J.}\ \bibnamefont {Schofield}}, \bibinfo {author} {\bibfnamefont {M.}~\bibnamefont {Chiao}}, \bibinfo {author} {\bibfnamefont {S.~R.}\ \bibnamefont {Julian}}, \bibinfo {author} {\bibfnamefont {G.~G.}\ \bibnamefont {Lonzarich}}, \bibinfo {author} {\bibfnamefont {S.~I.}\ \bibnamefont {Ikeda}}, \bibinfo {author} {\bibfnamefont {Y.}~\bibnamefont {Maeno}}, \bibinfo {author} {\bibfnamefont {A.~J.}\ \bibnamefont {Millis}},\ and\ \bibinfo {author} {\bibfnamefont {A.~P.}\ \bibnamefont {Mackenzie}},\ }\href {https://doi.org/10.1126/science.1063539} {\bibfield  {journal} {\bibinfo  {journal} {Science}\ }\textbf {\bibinfo {volume} {294}},\ \bibinfo {pages} {329} (\bibinfo {year} {2001})}\BibitemShut {NoStop}%
\bibitem [{\citenamefont {Grigera}\ \emph {et~al.}(2003)\citenamefont {Grigera}, \citenamefont {Borzi}, \citenamefont {Mackenzie}, \citenamefont {Julian}, \citenamefont {Perry},\ and\ \citenamefont {Maeno}}]{Sr3Ru2O7_angle}%
  \BibitemOpen
  \bibfield  {author} {\bibinfo {author} {\bibfnamefont {S.~A.}\ \bibnamefont {Grigera}}, \bibinfo {author} {\bibfnamefont {R.~A.}\ \bibnamefont {Borzi}}, \bibinfo {author} {\bibfnamefont {A.~P.}\ \bibnamefont {Mackenzie}}, \bibinfo {author} {\bibfnamefont {S.~R.}\ \bibnamefont {Julian}}, \bibinfo {author} {\bibfnamefont {R.~S.}\ \bibnamefont {Perry}},\ and\ \bibinfo {author} {\bibfnamefont {Y.}~\bibnamefont {Maeno}},\ }\href {https://doi.org/10.1103/PhysRevB.67.214427} {\bibfield  {journal} {\bibinfo  {journal} {Phys. Rev. B}\ }\textbf {\bibinfo {volume} {67}},\ \bibinfo {pages} {214427} (\bibinfo {year} {2003})}\BibitemShut {NoStop}%
\bibitem [{\citenamefont {Rosch}(1999)}]{disorder}%
  \BibitemOpen
  \bibfield  {author} {\bibinfo {author} {\bibfnamefont {A.}~\bibnamefont {Rosch}},\ }\href {https://doi.org/10.1103/PhysRevLett.82.4280} {\bibfield  {journal} {\bibinfo  {journal} {Phys. Rev. Lett.}\ }\textbf {\bibinfo {volume} {82}},\ \bibinfo {pages} {4280} (\bibinfo {year} {1999})}\BibitemShut {NoStop}%
\end{thebibliography}
\end{document}